\documentclass[nocopyrightspace,preprint]{sigplanconf} % and uncomment amspackages

\usepackage{lscape}
\usepackage{pstricks}
\usepackage{prooftree}

\usepackage{url}

\usepackage{isabelle,isabellesym} %general isabelle setup
\isabellestyle{it}

\newcommand{\tcomment}[1]{}

\newcommand{\isab}[1]{\begin{isabellebody} #1 \end{isabellebody}}

\newcommand{\fskip}[0] {\ensuremath{\;\;\;\;\;}}

\newcommand{\limp}[0] {\ensuremath{\rightarrow}}

\newcommand{\liff}[0] {\ensuremath{\leftrightarrow}}

\usepackage{amssymb,amsthm,amsmath}

\newtheorem{theorem}{Theorem}[section]
\newtheorem{corollary}[theorem]{Corollary}

\usepackage{amssymb}

\newcommand{\seq}[2]{#1 \vdash #2}

\newcommand{\Cut}{\ensuremath{\textit{Cut}}}
\newcommand{\Init}{\ensuremath{\textit{Init}}}

\newcommand{\craigRules}[0]{
\begin{center}
\noindent\begin{tabular*}{0.7\textwidth}{@{\extracolsep{\fill}}cc}

\begin{prooftree}
\justifies
{A, \Gamma \vdash \Delta, A}
\using
{\Init}
\end{prooftree}

\\ \\

\begin{prooftree}
\justifies
{\bot, \Gamma \vdash \Delta}
\using
{\bot L}
\end{prooftree}

&

\begin{prooftree}
\justifies
{\Gamma \vdash \Delta, \top}
\using
{\top R}
\end{prooftree}

\\ \\

\begin{prooftree}
{A, B, A \land B, \Gamma \vdash \Delta}
\justifies
{A \land B, \Gamma \vdash \Delta}
\using
{\land L}
\end{prooftree}

&

\begin{prooftree}
{\Gamma \vdash \Delta, A \land B, A} \fskip {\Gamma \vdash \Delta, A \land B, B}
\justifies
{\Gamma \vdash \Delta, A \land B}
\using
{\land R}
\end{prooftree}

\\ \\

\begin{prooftree}
{A, A \lor B, \Gamma \vdash \Delta} \fskip {B, A \lor B, \Gamma \vdash \Delta}
\justifies
{A \lor B, \Gamma \vdash \Delta}
\using
{\lor L}
\end{prooftree}

&

\begin{prooftree}
{\Gamma \vdash \Delta, A \lor B, A, B}
\justifies
{\Gamma \vdash \Delta, A \lor B}
\using
{\lor R}
\end{prooftree}

\\ \\

\begin{prooftree}
{\lnot A, \Gamma \vdash \Delta, A}
\justifies
{\lnot A, \Gamma \vdash \Delta}
\using
{\lnot L}
\end{prooftree}

&

\begin{prooftree}
{A, \Gamma \vdash \Delta, \lnot A}
\justifies
{\Gamma \vdash \Delta, \lnot A}
\using
{\lnot R}
\end{prooftree}

\\ \\

\begin{prooftree}
{A[t], \forall x. A[x], \Gamma \vdash \Delta}
\justifies
{\forall x. A[x], \Gamma \vdash \Delta}
\using
{\forall L}
\end{prooftree}

&

\begin{prooftree}
{\Gamma \vdash \Delta, \forall x. A[x], A[a]}
\justifies
{\Gamma \vdash \Delta, \forall x. A[x]}
\using
{\forall R}
\end{prooftree}

\\ \\

\begin{prooftree}
{A[a], \exists x. A[x], \Gamma \vdash \Delta}
\justifies
{\exists x. A[x], \Gamma \vdash \Delta}
\using
{\exists L}
\end{prooftree}

&

\begin{prooftree}
{\Gamma \vdash \Delta, \exists x. A[x], A[t]}
\justifies
{\Gamma \vdash \Delta, \exists x. A[x]}
\using
{\exists R}
\end{prooftree}

\\ \\

\begin{prooftree}
{\Gamma \vdash \Delta}
\justifies
{A, \Gamma \vdash \Delta}
\using
\WeakL
\end{prooftree}

&

\begin{prooftree}
{\Gamma \vdash \Delta}
\justifies
{\Gamma \vdash \Delta, A}
\using
\WeakR
\end{prooftree}

\end{tabular*}
\end{center}
}

\newcommand{\tsection}[1]{\section{#1}}
\newcommand{\tsubsection}[1]{\subsection{#1}}

\begin{document}

%\tcomment{
% sigplanconf info
\authorinfo{Tom Ridge}{University of Cambridge}{tjr22@cl.cam.ac.uk}
\title{Craig's Interpolation Theorem formalised and mechanised in Isabelle/HOL}
%\preprintfooter{Draft of \today.}
%\titlebanner{DRAFT of 2006-07-11.}
\maketitle
%}

\begin{abstract}
  
  We formalise and mechanise a construtive, proof theoretic proof of
  Craig's Interpolation Theorem in Isabelle/HOL. We give all the
  definitions and lemma statements both formally and informally. We
  also transcribe informally the formal proofs. We detail the main
  features of our mechanisation, such as the formalisation of binding
  for first order formulae. We also give some applications of
  Craig's Interpolation Theorem.

\end{abstract}

%\tableofcontents

%FIXME add motivation- formalised proof theory in the style of poplmark

%FIXME use of UNION- change to union of image of a function on a set?

%FIXME state explicitly poplmark stance- that binding is really not that difficult and that with careful phrasing you can avoid biasing the main mechanisation towards any particular implementation of binding

% FIXME notation section needs updating

\newcommand{\Weak}{\ensuremath{W}}
\newcommand{\WeakL}{{WL}}%{{\textit{Weak} L}}
\newcommand{\WeakR}{{WR}}%{{\textit{Weak} R}}

\tsection{Introduction}

\tcomment{
what, importance
why mechanise
how mechanise
features of mechanisation
overview
}

Craig's Interpolation Theorem is one of the main results in elementary
proof theory.  It is a result about FOL. Its proof is similar in style
to the more famous {\Cut} elimination theorem of Gentzen
\cite{szabo69gentzen}. In fact, the two results are intimately
connected, and both are part of a general concern with ``purity of
methods'' \cite{girard87proof}.

As with {\Cut} elimination, Craig's Interpolation Theorem has many
applications, particularly to the formalisation and mechanisation of
mathematics, to the making of definitions, to the stating of lemmas,
and to the general structuring of formalisations. It is primarily a
result about modularity at the level of definitions and lemmas.

This work describes the first mechanised proof of Craig's
Interpolation Theorem. Why mechanise Craig's Interpolation Theorem?
Correctness is one of the main considerations. Particularly, we would
like our proofs to be correctly formed (a purely syntactic condition),
even if we must use our own faculties to ensure the correctness of
definitions (that they conform to our informal notions).
Results in proof theory are particularly appropriate for formalisation
because they often involve substantial syntactic weight, which can
cause typographical and real errors to creep into non-mechanised
presentations.

A formal presentation also clarifies details, which in turn has
pedagogic advantages.
For example, the notion of variable binding and alpha conversion,
which are often viewed as tricky to establish formally,
are present in two places when formalising FOL. They are present when
considering variable binding $\forall x, \exists x$ in formulae. They
are also present in proof terms with the notion of an eigenvariable.
Much of the motivation behind the recent POPLmark challenge
\cite{poplmark} is to assess the current state of theorem provers with
regard to the mechanisation of proofs about logical systems,
particularly with respect to their handling of binding. There is
clearly a lot of interest in this area, and we believe our work
contains contributions. 

The proof of Craig's Interpolation Theorem we mechanise here is
constructive, which means that the proof contains an algorithm.  For a
given proof this algorithm constructs interpolation formulas. Thus,
the proof of the theorem is simultaneously the verification of an
algorithm. We believe this algorithm would be extremely hard to get
right without mechanical assistance, for exactly the same reasons that
it is hard to construct a correct informal proof: the details
overwhelm.

In this paper, we describe the result itself, and its mechanisation in
the Isabelle/HOL theorem prover. The mechanisation is presented in its
entirety, save that some tactic proof scripts have been omitted. The
paper should be readable with no Isabelle/HOL knowledge. By omitting
the Isabelle/HOL material, a standard informal mathematical
presentation is obtained. The full proof scripts can be obtained from
the author's homepage\footnote{\url{http://www.cl.cam.ac.uk/~tjr22/}}.

The mechanisation has several interesting features which we discuss
after the presentation of the main result.

We briefly outline the following sections. In Sect.
\ref{sectCraigNotation} we describe the formal syntax of Isabelle/HOL.
In Sect. \ref{sectCraigTerms} we describe terms, and in Sect.
\ref{sectCraigFormulae} we describe formulae. In Sect.
\ref{sectCraigSequents} we describe the system of FOL for which we
prove Craig's Interpolation Theorem. In Sect. \ref{sectCraigStatement}
we motivate the statement of (a strong form of) Craig's Interpolation
Theorem, and in Sect. \ref{sectCraigProof} we prove the theorem by
induction over derivations. Throughout we give both an informal
presentation, and the formal version for comparison.
Our development is axiomatic. To ensure that the axioms are
satisfiable, we also provide in Sect. \ref{sectCraigConcrete} a
concrete development which is conservative over the base Isabelle/HOL
logic.
In Sect. \ref{sectCraigAnalysis} we briefly analyse the mechanisation,
and then in Sect. \ref{sectCraigApplications} we discuss applications
of the theorem and its mechanisation.
Finally, we conclude with a statement of the main contributions of
this work, an examination of related work, and possibilities to extend
this work in the future.

%%%%%%%%%%%%%%%%%%%%%%%%%%%%%%%%%%%%%%%%%%%%%%%%%%%%%%%%%%%%%%%%%%%%%%%%%%%%%%%%
\tsection{Isabelle/HOL Notation}

\label{sectCraigNotation}

In the following sections, formal results are stated in the
Isabelle/HOL \cite{isabelle-dist} dialect of the HOL 
%cite{FIXME} 
logic.
\isab{\isamarkupfalse%
\begin{isamarkuptext}%
New types are introduced with the keyword \isacommand{typedecl}. New
  names for existing types (type aliases) are introduced with the
  keyword \isacommand{types}. 
  Type constructors are functions mapping type lists to types.
  Application of a type constructor is typically written postfix. For
  example, the type of sets over an underlying type \isa{{\isacharprime}a} is
  \isa{{\isacharprime}a\ set}. 
  The type of a function with domain \isa{{\isacharprime}a} and codomain \isa{{\isacharprime}b} is \isa{{\isacharprime}a\ {\isasymRightarrow}\ {\isacharprime}b}. \isa{{\isasymRightarrow}} is an infix type
  constructor, which associates to the right. Lambda abstraction $\lambda x$ is written \isa{{\isasymlambda}\ x}. 
  The type of pairs whose first component is of type
  \isa{{\isacharprime}a} and whose second component is of type \isa{{\isacharprime}b} is
  \isa{{\isacharprime}a\ {\isasymtimes}\ {\isacharprime}b}. The pair of $x$ and $y$ is written \isa{{\isacharparenleft}x{\isacharcomma}\ y{\isacharparenright}}.
  The type of lists whose elements are of type \isa{{\isacharprime}a} is \isa{{\isacharprime}a\ list}. Finite lists are written \isa{{\isacharbrackleft}a{\isacharcomma}b{\isacharcomma}c{\isacharbrackright}}. Consing an
  element \isa{x} onto the front of the list \isa{xs} is written
  \isa{x{\isacharhash}xs}.

A particularly important type is \isa{nat}, the type of the natural numbers. Non-recursive natural number elimination, or case analysis, is written \isa{case\ n\ of\ {\isadigit{0}}\ {\isasymRightarrow}\ a\ {\isasymor}\ Suc\ n{\isacharprime}\ {\isasymRightarrow}\ f\ n{\isacharprime}}.

  New constants are introduced with the keyword \isacommand{consts}. A
  new constant is introduced by giving its name followed by \isa{{\isacharcolon}{\isacharcolon}} followed by its type. Definitions are introduced with the keyword \isacommand{defs}. A definition is written using the metaequality \isa{{\isasymequiv}} rather than simple HOL equality. These two keywords are combined into the single keyword \isacommand{constdefs}.

  Axioms are introduced with the keyword \isacommand{axioms}.
  
  Our Isabelle theory files are ASCII text files. The format of these
  files is described in \cite{isabelle-dist}. The usual logical connectives
  are rendered in ASCII as follows. $\forall x. P\ x$ is \isa{{\isasymforall}\ x{\isachardot}\ P\ x}, $\exists x. P\ x$
  is \isa{{\isasymexists}\ x{\isachardot}\ P\ x}, $A \limp B$ is \isa{A\ {\isasymlongrightarrow}\ B}, $A \land B$ is \isa{A\ {\isasymand}\ B}, $A \lor B$
  is \isa{A\ {\isasymor}\ B}. $\lnot A$ is \isa{{\isachartilde}\ A}. $A \liff B$ is \isa{A\ {\isacharequal}\ B}.
  
  A common language is that of sets. Set notation is as follows. $a \in A$
  is \isa{a\ {\isasymin}\ A}. $a \not\in A$ is \isa{a\ {\isasymnotin}\ A}. The empty set is \isa{{\isacharbraceleft}{\isacharbraceright}}. Set union $A \cup B$ is \isa{A\ {\isasymunion}\ B}. Set intersection, $A \cap B$
  is \isa{A\ {\isasyminter}\ B}. Finite sets are written \isa{{\isacharbraceleft}a{\isacharcomma}b{\isacharcomma}c{\isacharbraceright}}. $A \subseteq B$ 
  is \isa{A\ {\isasymsubseteq}\ B}. The collection of the image of a function $f$ on a
  set $S$ is \isa{UNION\ S\ f}.

%let notation
  
  ML-style datatypes are introduced with the keyword
  \isacommand{datatype}, followed by the name of the new type,
  followed by constructors with the types of their arguments. The
  associated initial free structure with these constructors is then
  generated, together with various theorems about the structure.
  Functions can be defined by primitive recursion over the datatype.
  Primitive recursive functions are introduced with the keyword
  \isacommand{primrec}.%
\end{isamarkuptext}%
}
%%%%%%%%%%%%%%%%%%%%%%%%%%%%%%%%%%%%%%%%%%%%%%%%%%%%%%%%%%%%%%%%%%%%%%%%%%%%%%%%

\tsection{Terms}

\label{sectCraigTerms}

Variables are indexed by $\mathbb{N}$.

\isab{%
\endisatagtheory
{\isafoldtheory}%
\isadelimtheory
\isanewline
\endisadelimtheory
\isacommand{types}\isamarkupfalse%
\ var\ {\isacharequal}\ nat\isanewline
}
Terms are simply variables. 

\isab{\isanewline
\isacommand{types}\isamarkupfalse%
\ tm\ {\isacharequal}\ var\isanewline
}
The extension to full first order terms is
trivial. However, this obscures the development. Moreover, first order
terms can be simulated using variables and relations. 

%%%%%%%%%%%%%%%%%%%%%%%%%%%%%%%%%%%%%%%%%%%%%%%%%%%%%%%%%%%%%%%%%%%%%%%%%%%%%%%%
\tsection{Formulae, Occurrences}

\label{sectCraigFormulae}

Primitive formulae $P(x,y,z)$ are predicates $P$ applied to a tuple of
variables $(x,y,z)$. Predicates $P, Q, \ldots$ are in reality
identified by an index $i \in \mathbb{N}$, so that primitive
predicates are $P_0, P_1, \ldots$. Arbitrary length tuples
$(x,y,\ldots,z)$ are represented by lists.
Formulae $A$ are defined inductively in the usual way from primitive formulae using additional constructors $\bot, \top, \land, \lor, \lnot, \forall, \exists$. 

\isab{
\isanewline
\isacommand{types}\isamarkupfalse%
\ pred\ {\isacharequal}\ nat\isanewline
\isanewline
\isacommand{typedecl}\isamarkupfalse%
\ form\isanewline
\isanewline
\isacommand{consts}\isamarkupfalse%
\ \isanewline
\ \ P\ {\isacharcolon}{\isacharcolon}\ {\isachardoublequoteopen}pred\ {\isasymRightarrow}\ tm\ list\ {\isasymRightarrow}\ form{\isachardoublequoteclose}\isanewline
\ \ \ensuremath{\underline{\isasymbottom}}\ {\isacharcolon}{\isacharcolon}\ form\isanewline
\ \ \ensuremath{\underline{\isasymtop}}\ {\isacharcolon}{\isacharcolon}\ form\isanewline
\ \ \ensuremath{\underline{\isasymand}}\ {\isacharcolon}{\isacharcolon}\ {\isachardoublequoteopen}form\ {\isasymRightarrow}\ form\ {\isasymRightarrow}\ form{\isachardoublequoteclose}\isanewline
\ \ \ensuremath{\underline{\isasymor}}\ {\isacharcolon}{\isacharcolon}\ {\isachardoublequoteopen}form\ {\isasymRightarrow}\ form\ {\isasymRightarrow}\ form{\isachardoublequoteclose}\isanewline
\ \ \ensuremath{\underline{\isasymnot}}\ {\isacharcolon}{\isacharcolon}\ {\isachardoublequoteopen}form\ {\isasymRightarrow}\ form{\isachardoublequoteclose}\isanewline
\ \ \ensuremath{\underline{\isasymforall}}\ {\isacharcolon}{\isacharcolon}\ {\isachardoublequoteopen}var\ {\isasymRightarrow}\ form\ {\isasymRightarrow}\ form{\isachardoublequoteclose}\isanewline
\ \ \ensuremath{\underline{\isasymexists}}\ {\isacharcolon}{\isacharcolon}\ {\isachardoublequoteopen}var\ {\isasymRightarrow}\ form\ {\isasymRightarrow}\ form{\isachardoublequoteclose}\isanewline
}

Note that $P_i(x,y)$ is different to $P_i(x,y,z)$ but that later
definitions, such as \textit{pos,neg}, do not distinguish them. The
usual informal solution is not to work with two predicates of the same
name (index) but different arities. Alternatively a predicate could be
distinguished not only by its index, but also by its arity.

Informally, we often write quantified formula as $\forall x. A[x], \exists x.
A[x]$, and instantiations as $A[t], A[a]$.
The square brackets in quantified formulae $\forall x. A[x], \exists
x. A[x]$ and instantiations $A[t], A[a]$ have no formal meaning but
are intended to suggest the presence of occurrences in the body.
Sometimes they are intended to capture all occurrences in the body of
the formula, for instance, when writing $\forall x. A[x]$, we are
usually talking about all occurrences of $x$ in $A$. Other times they
are intended to capture only some occurrences in the body of the
formulae, for instance, when instantiating $\forall x. A[x]$ with a
term $t$, we write $A[t]$ to emphasise that the occurrences of $x$
have been replaced by $t$, even though $t$ may occur already in $A$.
Generally, $[]$ occurs in a rule that deals with binding. Then if $[]$
surrounds a bound variable, it matches all occurrences of the variable
in the term. If $[]$ surrounds a non-(bound variable) it binds some
occurrences (those where the bound variable previously appears).

$\forall$ and $\exists$ bind variables in the body of the formula.
We introduce auxiliary functions to handle instantiating quantifiers.
For example, $\textit{FAll\_inst}$ applied to a formula $\forall x.
A[x]$ and a term $t$ should produce the formula $A[t]$.

\isab{
\isanewline
\isacommand{consts}\isamarkupfalse%
\isanewline
\ \ FAll{\isacharunderscore}inst\ {\isacharcolon}{\isacharcolon}\ {\isachardoublequoteopen}tm\ {\isasymRightarrow}\ form\ {\isasymRightarrow}\ form{\isachardoublequoteclose}\isanewline
\ \ FEx{\isacharunderscore}inst\ {\isacharcolon}{\isacharcolon}\ {\isachardoublequoteopen}tm\ {\isasymRightarrow}\ form\ {\isasymRightarrow}\ form{\isachardoublequoteclose}\isanewline

\isanewline
\isacommand{axioms}\isamarkupfalse%
\isanewline
\ \ {\isachardoublequoteopen}{\isacharparenleft}FAll{\isacharunderscore}inst\ a\ {\isacharparenleft}\ensuremath{\underline{\isasymforall}}\ a\ C{\isacharparenright}{\isacharparenright}\ {\isacharequal}\ C{\isachardoublequoteclose}\ \isanewline
\ \ {\isachardoublequoteopen}{\isacharparenleft}FEx{\isacharunderscore}inst\ a\ {\isacharparenleft}\ensuremath{\underline{\isasymexists}}\ a\ C{\isacharparenright}{\isacharparenright}\ {\isacharequal}\ C{\isachardoublequoteclose}\ \isanewline
}

The free variables of a formula are defined as usual.

\isab{
\isanewline
\isacommand{consts}\isamarkupfalse%
\ fv\ {\isacharcolon}{\isacharcolon}\ {\isachardoublequoteopen}form\ {\isasymRightarrow}\ var\ list{\isachardoublequoteclose}\isanewline
\isanewline
\isacommand{axioms}\isamarkupfalse%
\isanewline
\ \ {\isachardoublequoteopen}a\ {\isasymnotin}\ {\isacharparenleft}set\ o\ fv{\isacharparenright}\ {\isacharparenleft}\ensuremath{\underline{\isasymforall}}\ a\ C{\isacharparenright}{\isachardoublequoteclose}\ \isanewline
\ \ {\isachardoublequoteopen}a\ {\isasymnotin}\ {\isacharparenleft}set\ o\ fv{\isacharparenright}\ {\isacharparenleft}\ensuremath{\underline{\isasymexists}}\ a\ C{\isacharparenright}{\isachardoublequoteclose}\ \isanewline
}

Positive and negative occurrences in a formula are defined.

\newcommand{\twocol}[2]{
\noindent\begin{minipage}[t]{0.5\textwidth}
#1
\end{minipage}
\hfill
\begin{minipage}[t]{0.5\textwidth}
#2
\end{minipage}
}

%\twocol
{\isab{
\isanewline
\isacommand{consts}\isamarkupfalse%
\ pos\ {\isacharcolon}{\isacharcolon}\ {\isachardoublequoteopen}form\ {\isasymRightarrow}\ pred\ set{\isachardoublequoteclose}\isanewline
\isanewline
\isacommand{axioms}\isamarkupfalse%
\ \ \isanewline
\ \ {\isachardoublequoteopen}pos\ {\isacharparenleft}P\ i\ tms{\isacharparenright}\ {\isacharequal}\ {\isacharbraceleft}i{\isacharbraceright}{\isachardoublequoteclose}\isanewline
\ \ {\isachardoublequoteopen}pos\ \ensuremath{\underline{\isasymbottom}}\ {\isacharequal}\ {\isacharbraceleft}{\isacharbraceright}{\isachardoublequoteclose}\isanewline
\ \ {\isachardoublequoteopen}pos\ \ensuremath{\underline{\isasymtop}}\ {\isacharequal}\ {\isacharbraceleft}{\isacharbraceright}{\isachardoublequoteclose}\isanewline
\ \ {\isachardoublequoteopen}pos\ {\isacharparenleft}\ensuremath{\underline{\isasymand}}\ A\ B{\isacharparenright}\ {\isacharequal}\ {\isacharparenleft}pos\ A{\isacharparenright}\ {\isasymunion}\ {\isacharparenleft}pos\ B{\isacharparenright}{\isachardoublequoteclose}\isanewline
\ \ {\isachardoublequoteopen}pos\ {\isacharparenleft}\ensuremath{\underline{\isasymor}}\ A\ B{\isacharparenright}\ {\isacharequal}\ {\isacharparenleft}pos\ A{\isacharparenright}\ {\isasymunion}\ {\isacharparenleft}pos\ B{\isacharparenright}{\isachardoublequoteclose}\isanewline
\ \ {\isachardoublequoteopen}pos\ {\isacharparenleft}\ensuremath{\underline{\isasymnot}}\ A{\isacharparenright}\ {\isacharequal}\ neg\ A{\isachardoublequoteclose}\isanewline
\ \ {\isachardoublequoteopen}pos\ {\isacharparenleft}\ensuremath{\underline{\isasymforall}}\ a\ A{\isacharparenright}\ {\isacharequal}\ pos\ A{\isachardoublequoteclose}\ \isanewline
\ \ {\isachardoublequoteopen}pos\ {\isacharparenleft}\ensuremath{\underline{\isasymexists}}\ a\ A{\isacharparenright}\ {\isacharequal}\ pos\ A{\isachardoublequoteclose}\ \isanewline
}}
{\isab{
\isanewline
\isacommand{consts}\isamarkupfalse%
\ neg\ {\isacharcolon}{\isacharcolon}\ {\isachardoublequoteopen}form\ {\isasymRightarrow}\ pred\ set{\isachardoublequoteclose}\isanewline
\isanewline
\isacommand{axioms}\isamarkupfalse%
\isanewline
\ \ {\isachardoublequoteopen}neg\ {\isacharparenleft}P\ i\ tms{\isacharparenright}\ {\isacharequal}\ {\isacharbraceleft}{\isacharbraceright}{\isachardoublequoteclose}\isanewline
\ \ {\isachardoublequoteopen}neg\ \ensuremath{\underline{\isasymbottom}}\ {\isacharequal}\ {\isacharbraceleft}{\isacharbraceright}{\isachardoublequoteclose}\isanewline
\ \ {\isachardoublequoteopen}neg\ \ensuremath{\underline{\isasymtop}}\ {\isacharequal}\ {\isacharbraceleft}{\isacharbraceright}{\isachardoublequoteclose}\isanewline
\ \ {\isachardoublequoteopen}neg\ {\isacharparenleft}\ensuremath{\underline{\isasymand}}\ A\ B{\isacharparenright}\ {\isacharequal}\ {\isacharparenleft}neg\ A{\isacharparenright}\ {\isasymunion}\ {\isacharparenleft}neg\ B{\isacharparenright}{\isachardoublequoteclose}\isanewline
\ \ {\isachardoublequoteopen}neg\ {\isacharparenleft}\ensuremath{\underline{\isasymor}}\ A\ B{\isacharparenright}\ {\isacharequal}\ {\isacharparenleft}neg\ A{\isacharparenright}\ {\isasymunion}\ {\isacharparenleft}neg\ B{\isacharparenright}{\isachardoublequoteclose}\isanewline
\ \ {\isachardoublequoteopen}neg\ {\isacharparenleft}\ensuremath{\underline{\isasymnot}}\ A{\isacharparenright}\ {\isacharequal}\ pos\ A{\isachardoublequoteclose}\isanewline
\ \ {\isachardoublequoteopen}neg\ {\isacharparenleft}\ensuremath{\underline{\isasymforall}}\ a\ A{\isacharparenright}\ {\isacharequal}\ neg\ A{\isachardoublequoteclose}\ \isanewline
\ \ {\isachardoublequoteopen}neg\ {\isacharparenleft}\ensuremath{\underline{\isasymexists}}\ a\ A{\isacharparenright}\ {\isacharequal}\ neg\ A{\isachardoublequoteclose}\ \isanewline
}}

\isab{\isanewline
\isacommand{axioms}\isamarkupfalse%
\isanewline
\ \ {\isachardoublequoteopen}pos\ {\isacharparenleft}FAll{\isacharunderscore}inst\ t\ {\isacharparenleft}\ensuremath{\underline{\isasymforall}}\ a\ A{\isacharparenright}{\isacharparenright}\ {\isacharequal}\ pos\ {\isacharparenleft}\ensuremath{\underline{\isasymforall}}\ a\ A{\isacharparenright}{\isachardoublequoteclose}\ \isanewline
\ \ {\isachardoublequoteopen}neg\ {\isacharparenleft}FAll{\isacharunderscore}inst\ t\ {\isacharparenleft}\ensuremath{\underline{\isasymforall}}\ a\ A{\isacharparenright}{\isacharparenright}\ {\isacharequal}\ neg\ {\isacharparenleft}\ensuremath{\underline{\isasymforall}}\ a\ A{\isacharparenright}{\isachardoublequoteclose}\ \isanewline
\isanewline
\ \ {\isachardoublequoteopen}pos\ {\isacharparenleft}FEx{\isacharunderscore}inst\ t\ {\isacharparenleft}\ensuremath{\underline{\isasymexists}}\ a\ A{\isacharparenright}{\isacharparenright}\ {\isacharequal}\ pos\ {\isacharparenleft}\ensuremath{\underline{\isasymexists}}\ a\ A{\isacharparenright}{\isachardoublequoteclose}\ \isanewline
\ \ {\isachardoublequoteopen}neg\ {\isacharparenleft}FEx{\isacharunderscore}inst\ t\ {\isacharparenleft}\ensuremath{\underline{\isasymexists}}\ a\ A{\isacharparenright}{\isacharparenright}\ {\isacharequal}\ neg\ {\isacharparenleft}\ensuremath{\underline{\isasymexists}}\ a\ A{\isacharparenright}{\isachardoublequoteclose}\ \isanewline
}

%%%%%%%%%%%%%%%%%%%%%%%%%%%%%%%%%%%%%%%%%%%%%%%%%%%%%%%%%%%%%%%%%%%%%%%%%%%%%%%%
\tsection{Sequents, Logical System}

\label{sectCraigSequents}

Sequents $\Gamma \vdash \Delta$ are pairs of sets of formulae. 
\isab{\isanewline
\isacommand{types}\isamarkupfalse%
\ seq\ {\isacharequal}\ {\isachardoublequoteopen}form\ set\ {\isacharasterisk}\ form\ set{\isachardoublequoteclose}\isanewline
}
The sets are intended to be finite. We make the restriction to finite sets of formulas when we define derivations.
We write sets of formulae using $,$ to denote (non-disjoint) set union. Thus
$\Gamma_1, \Gamma_2 = \Gamma_1 \cup \Gamma_2$.

\begin{figure*}

\craigRules

\vspace{0.5cm}

\center $\forall R, \exists L$: $a$ not free in the conclusion of the rule.

\caption{Rules for a Multiple Conclusion Sequent Calculus}
\label{figCraigRules}

\end{figure*}

We employ a standard multiple conclusion sequent calculus, see Fig. \ref{figCraigRules}.
Formulae in the conclusion of a rule are retained in the premises.
Exchange does not apply because we are working with sets of formulae.
Similarly contraction.  Weakening is actually admissible, but we
include it as an explicit rule because it makes the proofs more
elegant. For the weakening rules, it is important to recognise that $A$ may appear in $\Gamma, \Delta$.

The logical system describes the construction of a derivation. A derivation is a tree where each node is an instance of a rule. 

\isab{\isanewline
\isacommand{datatype}\isamarkupfalse%
\ deriv\ {\isacharequal}\ Init\ seq\isanewline
\ \ {\isasymor}\ \ensuremath{\underline{\isasymbottom}}L\ seq\ \isanewline
\ \ {\isasymor}\ \ensuremath{\underline{\isasymtop}}R\ seq\isanewline
\ \ {\isasymor}\ \ensuremath{\underline{\isasymand}}L\ seq\ deriv\isanewline
\ \ {\isasymor}\ \ensuremath{\underline{\isasymand}}R\ seq\ deriv\ deriv\isanewline
\ \ {\isasymor}\ \ensuremath{\underline{\isasymor}}L\ seq\ deriv\ deriv\isanewline
\ \ {\isasymor}\ \ensuremath{\underline{\isasymor}}R\ seq\ deriv\isanewline
\ \ {\isasymor}\ \ensuremath{\underline{\isasymnot}}L\ seq\ deriv\isanewline
\ \ {\isasymor}\ \ensuremath{\underline{\isasymnot}}R\ seq\ deriv\ \isanewline
\ \ {\isasymor}\ \ensuremath{\underline{{\isasymforall}}}L\ seq\ deriv\isanewline
\ \ {\isasymor}\ \ensuremath{\underline{{\isasymforall}}}R\ seq\ deriv\isanewline
\ \ {\isasymor}\ \ensuremath{\underline{{\isasymexists}}}L\ seq\ deriv\isanewline
\ \ {\isasymor}\ \ensuremath{\underline{{\isasymexists}}}R\ seq\ deriv\isanewline
\ \ {\isasymor}\ WL\ seq\ deriv\isanewline
\ \ {\isasymor}\ WR\ seq\ deriv\isanewline
}
The first argument to each derivation constructor indicates the root
sequent of the derivation formed using the constructor. The additional
arguments provide auxiliary information necessary to determine the
rule. For example, in the case of $\land L$, we must give the formulas
$A$ and $B$ where $A \land B$ is the formula we are analysing, and we
must also provide a subderivation of the premise of the rule. The exact requirements are explicitly stated when we define \textit{is-deriv}.
%
%For Example, The Following Derivation

%Fixme

%Would Be Represented By The Term

%Fixme

The root of a derivation is straightforward.

\isab{\isanewline
\isacommand{consts}\isamarkupfalse%
\ root\ {\isacharcolon}{\isacharcolon}\ {\isachardoublequoteopen}deriv\ {\isasymRightarrow}\ seq{\isachardoublequoteclose}\isanewline
\isacommand{primrec}\isamarkupfalse%
\ \isanewline
\ \ {\isachardoublequoteopen}root\ {\isacharparenleft}Init\ {\isasymGamma}{\isasymDelta}{\isacharparenright}\ {\isacharequal}\ {\isasymGamma}{\isasymDelta}{\isachardoublequoteclose}\isanewline
\ \ {\isachardoublequoteopen}root\ {\isacharparenleft}\ensuremath{\underline{\isasymbottom}}L\ {\isasymGamma}{\isasymDelta}{\isacharparenright}\ {\isacharequal}\ {\isasymGamma}{\isasymDelta}{\isachardoublequoteclose}\isanewline
\ \ {\isachardoublequoteopen}root\ {\isacharparenleft}\ensuremath{\underline{\isasymtop}}R\ {\isasymGamma}{\isasymDelta}{\isacharparenright}\ {\isacharequal}\ {\isasymGamma}{\isasymDelta}{\isachardoublequoteclose}\isanewline
\ \ {\isachardoublequoteopen}root\ {\isacharparenleft}\ensuremath{\underline{\isasymand}}L\ {\isasymGamma}{\isasymDelta}\ d{\isacharparenright}\ {\isacharequal}\ {\isasymGamma}{\isasymDelta}{\isachardoublequoteclose}\isanewline
\ \ {\isachardoublequoteopen}root\ {\isacharparenleft}\ensuremath{\underline{\isasymand}}R\ {\isasymGamma}{\isasymDelta}\ dl\ dr{\isacharparenright}\ {\isacharequal}\ {\isasymGamma}{\isasymDelta}{\isachardoublequoteclose}\isanewline
\ \ {\isachardoublequoteopen}root\ {\isacharparenleft}\ensuremath{\underline{\isasymor}}L\ {\isasymGamma}{\isasymDelta}\ dl\ dr{\isacharparenright}\ {\isacharequal}\ {\isasymGamma}{\isasymDelta}{\isachardoublequoteclose}\isanewline
\ \ {\isachardoublequoteopen}root\ {\isacharparenleft}\ensuremath{\underline{\isasymor}}R\ {\isasymGamma}{\isasymDelta}\ d{\isacharparenright}\ {\isacharequal}\ {\isasymGamma}{\isasymDelta}{\isachardoublequoteclose}\isanewline
\ \ {\isachardoublequoteopen}root\ {\isacharparenleft}\ensuremath{\underline{\isasymnot}}L\ {\isasymGamma}{\isasymDelta}\ d{\isacharparenright}\ {\isacharequal}\ {\isasymGamma}{\isasymDelta}{\isachardoublequoteclose}\isanewline
\ \ {\isachardoublequoteopen}root\ {\isacharparenleft}\ensuremath{\underline{\isasymnot}}R\ {\isasymGamma}{\isasymDelta}\ d{\isacharparenright}\ {\isacharequal}\ {\isasymGamma}{\isasymDelta}{\isachardoublequoteclose}\isanewline
\ \ {\isachardoublequoteopen}root\ {\isacharparenleft}\ensuremath{\underline{{\isasymforall}}}L\ {\isasymGamma}{\isasymDelta}\ d{\isacharparenright}\ {\isacharequal}\ {\isasymGamma}{\isasymDelta}{\isachardoublequoteclose}\isanewline
\ \ {\isachardoublequoteopen}root\ {\isacharparenleft}\ensuremath{\underline{{\isasymforall}}}R\ {\isasymGamma}{\isasymDelta}\ d{\isacharparenright}\ {\isacharequal}\ {\isasymGamma}{\isasymDelta}{\isachardoublequoteclose}\isanewline
\ \ {\isachardoublequoteopen}root\ {\isacharparenleft}\ensuremath{\underline{{\isasymexists}}}L\ {\isasymGamma}{\isasymDelta}\ d{\isacharparenright}\ {\isacharequal}\ {\isasymGamma}{\isasymDelta}{\isachardoublequoteclose}\isanewline
\ \ {\isachardoublequoteopen}root\ {\isacharparenleft}\ensuremath{\underline{{\isasymexists}}}R\ {\isasymGamma}{\isasymDelta}\ d{\isacharparenright}\ {\isacharequal}\ {\isasymGamma}{\isasymDelta}{\isachardoublequoteclose}\isanewline
\ \ {\isachardoublequoteopen}root\ {\isacharparenleft}WL\ {\isasymGamma}{\isasymDelta}\ d{\isacharparenright}\ {\isacharequal}\ {\isasymGamma}{\isasymDelta}{\isachardoublequoteclose}\isanewline
\ \ {\isachardoublequoteopen}root\ {\isacharparenleft}WR\ {\isasymGamma}{\isasymDelta}\ d{\isacharparenright}\ {\isacharequal}\ {\isasymGamma}{\isasymDelta}{\isachardoublequoteclose}\isanewline
}
We use a predicate to pick out wellformed derivations.

%FIXME \renewcommand{\tsmall}{\scriptsize}
%\twocol
{
\isab{\isanewline
\isacommand{consts}\isamarkupfalse%
\ is{\isacharunderscore}deriv\ {\isacharcolon}{\isacharcolon}\ {\isachardoublequoteopen}deriv\ {\isasymRightarrow}\ bool{\isachardoublequoteclose}\isanewline
\isacommand{primrec}\isamarkupfalse%
\isanewline
\ \ {\isachardoublequoteopen}is{\isacharunderscore}deriv\ {\isacharparenleft}Init\ {\isasymGamma}{\isasymDelta}{\isacharparenright}\ {\isacharequal}\ {\isacharparenleft}let\ {\isacharparenleft}{\isasymGamma}{\isacharcomma}{\isasymDelta}{\isacharparenright}\ {\isacharequal}\ {\isasymGamma}{\isasymDelta}\ in\ \isanewline
\ \ \ \ finite\ {\isasymGamma}\ \isanewline
\ \ \ \ {\isasymand}\ finite\ {\isasymDelta}\ \isanewline
\ \ \ \ {\isasymand}\ {\isacharparenleft}{\isasymexists}\ A{\isachardot}\ A\ {\isasymin}\ {\isasymGamma}\ \isanewline
\ \ \ \ {\isasymand}\ A\ {\isasymin}\ {\isasymDelta}{\isacharparenright}{\isacharparenright}{\isachardoublequoteclose}\isanewline
\isanewline
\ \ {\isachardoublequoteopen}is{\isacharunderscore}deriv\ {\isacharparenleft}\ensuremath{\underline{\isasymbottom}}L\ {\isasymGamma}{\isasymDelta}{\isacharparenright}\ {\isacharequal}\ {\isacharparenleft}let\ {\isacharparenleft}{\isasymGamma}{\isacharcomma}{\isasymDelta}{\isacharparenright}\ {\isacharequal}\ {\isasymGamma}{\isasymDelta}\ in\ \isanewline
\ \ \ \ finite\ {\isasymGamma}\ \isanewline
\ \ \ \ {\isasymand}\ finite\ {\isasymDelta}\ \isanewline
\ \ \ \ {\isasymand}\ \ensuremath{\underline{\isasymbottom}}\ {\isasymin}\ {\isasymGamma}{\isacharparenright}{\isachardoublequoteclose}\isanewline
\isanewline
\ \ {\isachardoublequoteopen}is{\isacharunderscore}deriv\ {\isacharparenleft}\ensuremath{\underline{\isasymtop}}R\ {\isasymGamma}{\isasymDelta}{\isacharparenright}\ {\isacharequal}\ {\isacharparenleft}let\ {\isacharparenleft}{\isasymGamma}{\isacharcomma}{\isasymDelta}{\isacharparenright}\ {\isacharequal}\ {\isasymGamma}{\isasymDelta}\ in\ \isanewline
\ \ \ \ finite\ {\isasymGamma}\ \isanewline
\ \ \ \ {\isasymand}\ finite\ {\isasymDelta}\ \isanewline
\ \ \ \ {\isasymand}\ \ensuremath{\underline{\isasymtop}}\ {\isasymin}\ {\isasymDelta}{\isacharparenright}{\isachardoublequoteclose}\isanewline
\isanewline
\ \ {\isachardoublequoteopen}is{\isacharunderscore}deriv\ {\isacharparenleft}\ensuremath{\underline{\isasymand}}L\ {\isasymGamma}{\isasymDelta}\ d{\isacharparenright}\ {\isacharequal}\ {\isacharparenleft}let\ {\isacharparenleft}{\isasymGamma}{\isacharcomma}{\isasymDelta}{\isacharparenright}\ {\isacharequal}\ {\isasymGamma}{\isasymDelta}\ in\ \isanewline
\ \ \ \ finite\ {\isasymGamma}\ \isanewline
\ \ \ \ {\isasymand}\ finite\ {\isasymDelta}\ \isanewline
\ \ \ \ {\isasymand}\ {\isacharparenleft}{\isasymexists}\ A\ B{\isachardot}\ \ensuremath{\underline{\isasymand}}\ A\ B\ {\isasymin}\ {\isasymGamma}\ \isanewline
\ \ \ \ {\isasymand}\ is{\isacharunderscore}deriv\ d\ \isanewline
\ \ \ \ {\isasymand}\ root\ d\ {\isacharequal}\ {\isacharparenleft}{\isacharbraceleft}A{\isacharcomma}B{\isacharbraceright}\ {\isasymunion}\ {\isasymGamma}{\isacharcomma}{\isasymDelta}{\isacharparenright}{\isacharparenright}{\isacharparenright}{\isachardoublequoteclose}\isanewline
\isanewline
\ \ {\isachardoublequoteopen}is{\isacharunderscore}deriv\ {\isacharparenleft}\ensuremath{\underline{\isasymand}}R\ {\isasymGamma}{\isasymDelta}\ dl\ dr{\isacharparenright}\ {\isacharequal}\ {\isacharparenleft}let\ {\isacharparenleft}{\isasymGamma}{\isacharcomma}{\isasymDelta}{\isacharparenright}\ {\isacharequal}\ {\isasymGamma}{\isasymDelta}\ in\ \isanewline
\ \ \ \ finite\ {\isasymGamma}\ \isanewline
\ \ \ \ {\isasymand}\ finite\ {\isasymDelta}\ \isanewline
\ \ \ \ {\isasymand}\ {\isacharparenleft}{\isasymexists}\ A\ B{\isachardot}\ \ensuremath{\underline{\isasymand}}\ A\ B\ {\isasymin}\ {\isasymDelta}\ \isanewline
\ \ \ \ {\isasymand}\ is{\isacharunderscore}deriv\ dl\ \isanewline
\ \ \ \ {\isasymand}\ root\ dl\ {\isacharequal}\ {\isacharparenleft}{\isasymGamma}{\isacharcomma}{\isasymDelta}\ {\isasymunion}\ {\isacharbraceleft}A{\isacharbraceright}{\isacharparenright}\ \isanewline
\ \ \ \ {\isasymand}\ is{\isacharunderscore}deriv\ dr\ \isanewline
\ \ \ \ {\isasymand}\ root\ dr\ {\isacharequal}\ {\isacharparenleft}{\isasymGamma}{\isacharcomma}{\isasymDelta}\ {\isasymunion}\ {\isacharbraceleft}B{\isacharbraceright}{\isacharparenright}{\isacharparenright}{\isacharparenright}{\isachardoublequoteclose}\isanewline
\isanewline
\ \ {\isachardoublequoteopen}is{\isacharunderscore}deriv\ {\isacharparenleft}\ensuremath{\underline{\isasymor}}L\ {\isasymGamma}{\isasymDelta}\ dl\ dr{\isacharparenright}\ {\isacharequal}\ {\isacharparenleft}let\ {\isacharparenleft}{\isasymGamma}{\isacharcomma}{\isasymDelta}{\isacharparenright}\ {\isacharequal}\ {\isasymGamma}{\isasymDelta}\ in\ \isanewline
\ \ \ \ finite\ {\isasymGamma}\ \isanewline
\ \ \ \ {\isasymand}\ finite\ {\isasymDelta}\ \isanewline
\ \ \ \ {\isasymand}\ {\isacharparenleft}{\isasymexists}\ A\ B{\isachardot}\ \ensuremath{\underline{\isasymor}}\ A\ B\ {\isasymin}\ {\isasymGamma}\ \isanewline
\ \ \ \ {\isasymand}\ is{\isacharunderscore}deriv\ dl\ \isanewline
\ \ \ \ {\isasymand}\ root\ dl\ {\isacharequal}\ {\isacharparenleft}{\isacharbraceleft}A{\isacharbraceright}\ {\isasymunion}\ {\isasymGamma}{\isacharcomma}{\isasymDelta}{\isacharparenright}\ \isanewline
\ \ \ \ {\isasymand}\ is{\isacharunderscore}deriv\ dr\ \isanewline
\ \ \ \ {\isasymand}\ root\ dr\ {\isacharequal}\ {\isacharparenleft}{\isacharbraceleft}B{\isacharbraceright}\ {\isasymunion}\ {\isasymGamma}{\isacharcomma}{\isasymDelta}{\isacharparenright}{\isacharparenright}{\isacharparenright}{\isachardoublequoteclose}\isanewline
\isanewline
\ \ {\isachardoublequoteopen}is{\isacharunderscore}deriv\ {\isacharparenleft}\ensuremath{\underline{\isasymor}}R\ {\isasymGamma}{\isasymDelta}\ d{\isacharparenright}\ {\isacharequal}\ {\isacharparenleft}let\ {\isacharparenleft}{\isasymGamma}{\isacharcomma}{\isasymDelta}{\isacharparenright}\ {\isacharequal}\ {\isasymGamma}{\isasymDelta}\ in\ \isanewline
\ \ \ \ finite\ {\isasymGamma}\ \isanewline
\ \ \ \ {\isasymand}\ finite\ {\isasymDelta}\ \isanewline
\ \ \ \ {\isasymand}\ {\isacharparenleft}{\isasymexists}\ A\ B{\isachardot}\ \ensuremath{\underline{\isasymor}}\ A\ B\ {\isasymin}\ {\isasymDelta}\ \isanewline
\ \ \ \ {\isasymand}\ is{\isacharunderscore}deriv\ d\ \isanewline
\ \ \ \ {\isasymand}\ root\ d\ {\isacharequal}\ {\isacharparenleft}{\isasymGamma}{\isacharcomma}{\isasymDelta}\ {\isasymunion}\ {\isacharbraceleft}A{\isacharcomma}B{\isacharbraceright}{\isacharparenright}{\isacharparenright}{\isacharparenright}{\isachardoublequoteclose}\isanewline
\isanewline
\ \ {\isachardoublequoteopen}is{\isacharunderscore}deriv\ {\isacharparenleft}\ensuremath{\underline{\isasymnot}}L\ {\isasymGamma}{\isasymDelta}\ d{\isacharparenright}\ {\isacharequal}\ {\isacharparenleft}let\ {\isacharparenleft}{\isasymGamma}{\isacharcomma}{\isasymDelta}{\isacharparenright}\ {\isacharequal}\ {\isasymGamma}{\isasymDelta}\ in\ \isanewline
\ \ \ \ finite\ {\isasymGamma}\ \isanewline
\ \ \ \ {\isasymand}\ finite\ {\isasymDelta}\ \isanewline
\ \ \ \ {\isasymand}\ {\isacharparenleft}{\isasymexists}\ C{\isachardot}\ \ensuremath{\underline{\isasymnot}}\ C\ {\isasymin}\ {\isasymGamma}\ \isanewline
\ \ \ \ {\isasymand}\ is{\isacharunderscore}deriv\ d\ \isanewline
\ \ \ \ {\isasymand}\ root\ d\ {\isacharequal}\ {\isacharparenleft}{\isasymGamma}{\isacharcomma}{\isasymDelta}\ {\isasymunion}\ {\isacharbraceleft}C{\isacharbraceright}{\isacharparenright}{\isacharparenright}{\isacharparenright}{\isachardoublequoteclose}\isanewline
}}{
\isab{\isanewline
\ \ {\isachardoublequoteopen}is{\isacharunderscore}deriv\ {\isacharparenleft}\ensuremath{\underline{\isasymnot}}R\ {\isasymGamma}{\isasymDelta}\ d{\isacharparenright}\ {\isacharequal}\ {\isacharparenleft}let\ {\isacharparenleft}{\isasymGamma}{\isacharcomma}{\isasymDelta}{\isacharparenright}\ {\isacharequal}\ {\isasymGamma}{\isasymDelta}\ in\ \isanewline
\ \ \ \ finite\ {\isasymGamma}\ \isanewline
\ \ \ \ {\isasymand}\ finite\ {\isasymDelta}\ \isanewline
\ \ \ \ {\isasymand}\ {\isacharparenleft}{\isasymexists}\ C{\isachardot}\ \ensuremath{\underline{\isasymnot}}\ C\ {\isasymin}\ {\isasymDelta}\ \isanewline
\ \ \ \ {\isasymand}\ is{\isacharunderscore}deriv\ d\ \isanewline
\ \ \ \ {\isasymand}\ root\ d\ {\isacharequal}\ {\isacharparenleft}{\isacharbraceleft}C{\isacharbraceright}\ {\isasymunion}\ {\isasymGamma}{\isacharcomma}{\isasymDelta}{\isacharparenright}{\isacharparenright}{\isacharparenright}{\isachardoublequoteclose}\isanewline
\isanewline
\ \ {\isachardoublequoteopen}is{\isacharunderscore}deriv\ {\isacharparenleft}\ensuremath{\underline{{\isasymforall}}}L\ {\isasymGamma}{\isasymDelta}\ d{\isacharparenright}\ {\isacharequal}\ {\isacharparenleft}\isanewline
\ \ \ \ let\ {\isacharparenleft}{\isasymGamma}{\isacharcomma}{\isasymDelta}{\isacharparenright}\ {\isacharequal}\ {\isasymGamma}{\isasymDelta}\ in\ \isanewline
\ \ \ \ finite\ {\isasymGamma}\ \isanewline
\ \ \ \ {\isasymand}\ finite\ {\isasymDelta}\ \isanewline
\ \ \ \ {\isasymand}\ {\isacharparenleft}{\isasymexists}\ A\ a\ t{\isachardot}\ \ensuremath{\underline{\isasymforall}}\ a\ A\ {\isasymin}\ {\isasymGamma}\ \isanewline
\ \ \ \ {\isasymand}\ is{\isacharunderscore}deriv\ d\ \isanewline
\ \ \ \ {\isasymand}\ root\ d\ {\isacharequal}\ {\isacharparenleft}{\isacharbraceleft}FAll{\isacharunderscore}inst\ t\ {\isacharparenleft}\ensuremath{\underline{\isasymforall}}\ a\ A{\isacharparenright}{\isacharbraceright}\ {\isasymunion}\ {\isasymGamma}{\isacharcomma}{\isasymDelta}{\isacharparenright}{\isacharparenright}{\isacharparenright}{\isachardoublequoteclose}\isanewline
\isanewline
\ \ {\isachardoublequoteopen}is{\isacharunderscore}deriv\ {\isacharparenleft}\ensuremath{\underline{{\isasymforall}}}R\ {\isasymGamma}{\isasymDelta}\ d{\isacharparenright}\ {\isacharequal}\ {\isacharparenleft}\isanewline
\ \ \ \ let\ {\isacharparenleft}{\isasymGamma}{\isacharcomma}{\isasymDelta}{\isacharparenright}\ {\isacharequal}\ {\isasymGamma}{\isasymDelta}\ in\ \isanewline
\ \ \ \ finite\ {\isasymGamma}\ \isanewline
\ \ \ \ {\isasymand}\ finite\ {\isasymDelta}\ \isanewline
\ \ \ \ {\isasymand}\ {\isacharparenleft}{\isasymexists}\ a\ A{\isachardot}\ \ensuremath{\underline{\isasymforall}}\ a\ A\ {\isasymin}\ {\isasymDelta}\ \isanewline
\ \ \ \ {\isasymand}\ a\ {\isasymnotin}\ UNION\ {\isacharparenleft}{\isasymGamma}\ {\isasymunion}\ {\isasymDelta}{\isacharparenright}\ {\isacharparenleft}set\ o\ fv{\isacharparenright}\isanewline
\ \ \ \ {\isasymand}\ is{\isacharunderscore}deriv\ d\ \isanewline
\ \ \ \ {\isasymand}\ root\ d\ {\isacharequal}\ {\isacharparenleft}{\isasymGamma}{\isacharcomma}{\isasymDelta}\ {\isasymunion}\ {\isacharbraceleft}A{\isacharbraceright}{\isacharparenright}{\isacharparenright}{\isacharparenright}{\isachardoublequoteclose}\isanewline
\isanewline
\ \ {\isachardoublequoteopen}is{\isacharunderscore}deriv\ {\isacharparenleft}\ensuremath{\underline{{\isasymexists}}}L\ {\isasymGamma}{\isasymDelta}\ d{\isacharparenright}\ {\isacharequal}\ {\isacharparenleft}\isanewline
\ \ \ \ let\ {\isacharparenleft}{\isasymGamma}{\isacharcomma}{\isasymDelta}{\isacharparenright}\ {\isacharequal}\ {\isasymGamma}{\isasymDelta}\ in\ \isanewline
\ \ \ \ finite\ {\isasymGamma}\ \isanewline
\ \ \ \ {\isasymand}\ finite\ {\isasymDelta}\ \isanewline
\ \ \ \ {\isasymand}\ {\isacharparenleft}{\isasymexists}\ a\ A{\isachardot}\ \ensuremath{\underline{\isasymexists}}\ a\ A\ {\isasymin}\ {\isasymGamma}\ \isanewline
\ \ \ \ {\isasymand}\ a\ {\isasymnotin}\ UNION\ {\isacharparenleft}{\isasymGamma}\ {\isasymunion}\ {\isasymDelta}{\isacharparenright}\ {\isacharparenleft}set\ o\ fv{\isacharparenright}\isanewline
\ \ \ \ {\isasymand}\ is{\isacharunderscore}deriv\ d\ \isanewline
\ \ \ \ {\isasymand}\ root\ d\ {\isacharequal}\ {\isacharparenleft}{\isacharbraceleft}A{\isacharbraceright}\ {\isasymunion}\ {\isasymGamma}{\isacharcomma}{\isasymDelta}{\isacharparenright}{\isacharparenright}{\isacharparenright}{\isachardoublequoteclose}\isanewline
\isanewline
\ \ {\isachardoublequoteopen}is{\isacharunderscore}deriv\ {\isacharparenleft}\ensuremath{\underline{{\isasymexists}}}R\ {\isasymGamma}{\isasymDelta}\ d{\isacharparenright}\ {\isacharequal}\ {\isacharparenleft}\isanewline
\ \ \ \ let\ {\isacharparenleft}{\isasymGamma}{\isacharcomma}{\isasymDelta}{\isacharparenright}\ {\isacharequal}\ {\isasymGamma}{\isasymDelta}\ in\ \isanewline
\ \ \ \ finite\ {\isasymGamma}\ \isanewline
\ \ \ \ {\isasymand}\ finite\ {\isasymDelta}\ \isanewline
\ \ \ \ {\isasymand}\ {\isacharparenleft}{\isasymexists}\ A\ a\ t{\isachardot}\ \ensuremath{\underline{\isasymexists}}\ a\ A\ {\isasymin}\ {\isasymDelta}\ \isanewline
\ \ \ \ {\isasymand}\ is{\isacharunderscore}deriv\ d\ \isanewline
\ \ \ \ {\isasymand}\ root\ d\ {\isacharequal}\ {\isacharparenleft}{\isasymGamma}{\isacharcomma}{\isasymDelta}\ {\isasymunion}\ {\isacharbraceleft}FEx{\isacharunderscore}inst\ t\ {\isacharparenleft}\ensuremath{\underline{\isasymexists}}\ a\ A{\isacharparenright}{\isacharbraceright}{\isacharparenright}{\isacharparenright}{\isacharparenright}{\isachardoublequoteclose}\isanewline
\isanewline
\ \ {\isachardoublequoteopen}is{\isacharunderscore}deriv\ {\isacharparenleft}WL\ {\isasymGamma}{\isasymDelta}\ d{\isacharparenright}\ {\isacharequal}\ {\isacharparenleft}{\isasymexists}\ {\isasymGamma}\ {\isasymDelta}\ A{\isachardot}\ \isanewline
\ \ \ \ finite\ {\isasymGamma}\ \isanewline
\ \ \ \ {\isasymand}\ finite\ {\isasymDelta}\ \isanewline
\ \ \ \ {\isasymand}\ is{\isacharunderscore}deriv\ d\ \isanewline
\ \ \ \ {\isasymand}\ root\ d\ {\isacharequal}\ {\isacharparenleft}{\isasymGamma}{\isacharcomma}{\isasymDelta}{\isacharparenright}\ \isanewline
\ \ \ \ {\isasymand}\ {\isasymGamma}{\isasymDelta}\ {\isacharequal}\ {\isacharparenleft}{\isacharbraceleft}A{\isacharbraceright}\ {\isasymunion}\ {\isasymGamma}{\isacharcomma}{\isasymDelta}{\isacharparenright}{\isacharparenright}{\isachardoublequoteclose}\isanewline
\isanewline
\ \ {\isachardoublequoteopen}is{\isacharunderscore}deriv\ {\isacharparenleft}WR\ {\isasymGamma}{\isasymDelta}\ d{\isacharparenright}\ {\isacharequal}\ {\isacharparenleft}{\isasymexists}\ {\isasymGamma}\ {\isasymDelta}\ A{\isachardot}\ \isanewline
\ \ \ \ finite\ {\isasymGamma}\ \isanewline
\ \ \ \ {\isasymand}\ finite\ {\isasymDelta}\ \isanewline
\ \ \ \ {\isasymand}\ is{\isacharunderscore}deriv\ d\ \isanewline
\ \ \ \ {\isasymand}\ root\ d\ {\isacharequal}\ {\isacharparenleft}{\isasymGamma}{\isacharcomma}{\isasymDelta}{\isacharparenright}\ \isanewline
\ \ \ \ {\isasymand}\ {\isasymGamma}{\isasymDelta}\ {\isacharequal}\ {\isacharparenleft}{\isasymGamma}{\isacharcomma}{\isasymDelta}\ {\isasymunion}\ {\isacharbraceleft}A{\isacharbraceright}{\isacharparenright}{\isacharparenright}{\isachardoublequoteclose}\isanewline
}}
%\renewcommand{\tsmall}{\small}

%%%%%%%%%%%%%%%%%%%%%%%%%%%%%%%%%%%%%%%%%%%%%%%%%%%%%%%%%%%%%%%%%%%%%%%%%%%%%%%%
\tsection{Statement of Craig's Interpolation Theorem}

\label{sectCraigStatement}

\newcommand{\skipand}[0]{\fskip \textit{and} \fskip}

\begin{theorem} (Craig's Interpolation Theorem) If 

\[
\seq{\Gamma}{\Delta}
\]

then there exists a formula $C$ such that

\[
\seq{\Gamma}{C} \skipand \seq{C}{\Delta}
\]

and moreover such that 

\begin{itemize}
\item Any predicate that occurs positively in $C$ occurs positively in $\Gamma$ \emph{and} in $\Delta$.
\item Any predicate that occurs negatively in $C$ occurs negatively in $\Gamma$ \emph{and} in $\Delta$.
\end{itemize}

\end{theorem}

\isab{%
\endisatagproof
{\isafoldproof}%
\isadelimproof
\isanewline
\endisadelimproof
\isacommand{lemma}\isamarkupfalse%
\ craig{\isacharcolon}\ {\isachardoublequoteopen}\isanewline
\ \ {\isasymforall}\ d\ {\isasymGamma}\ {\isasymDelta}{\isachardot}\isanewline
\ \ \ \ is{\isacharunderscore}deriv\ d\ \isanewline
\ \ \ \ {\isasymand}\ root\ d\ {\isacharequal}\ {\isacharparenleft}{\isasymGamma}{\isacharcomma}{\isasymDelta}{\isacharparenright}\ \isanewline
\ \ \ \ {\isasymlongrightarrow}\ \isanewline
\ \ \ \ {\isacharparenleft}{\isasymexists}\ C{\isachardot}\isanewline
\ \ \ \ \ \ {\isacharparenleft}{\isasymexists}\ dl{\isachardot}\ is{\isacharunderscore}deriv\ dl\ {\isasymand}\ root\ dl\ {\isacharequal}\ {\isacharparenleft}{\isasymGamma}{\isacharcomma}{\isacharbraceleft}C{\isacharbraceright}{\isacharparenright}{\isacharparenright}\isanewline
\ \ \ \ \ \ {\isasymand}\ {\isacharparenleft}{\isasymexists}\ dr{\isachardot}\ is{\isacharunderscore}deriv\ dr\ {\isasymand}\ root\ dr\ {\isacharequal}\ {\isacharparenleft}{\isacharbraceleft}C{\isacharbraceright}{\isacharcomma}{\isasymDelta}{\isacharparenright}{\isacharparenright}\isanewline
\ \ \ \ \ \ {\isasymand}\ {\isacharparenleft}pos\ C\ {\isasymsubseteq}\ {\isacharparenleft}UNION\ {\isasymGamma}\ pos{\isacharparenright}\ {\isasyminter}\ {\isacharparenleft}UNION\ {\isasymDelta}\ pos{\isacharparenright}{\isacharparenright}\isanewline
\ \ \ \ \ \ {\isasymand}\ {\isacharparenleft}neg\ C\ {\isasymsubseteq}\ {\isacharparenleft}UNION\ {\isasymGamma}\ neg{\isacharparenright}\ {\isasyminter}\ {\isacharparenleft}UNION\ {\isasymDelta}\ neg{\isacharparenright}{\isacharparenright}{\isacharparenright}\isanewline
{\isachardoublequoteclose}\isanewline
}
Craig's interpolation theorem is almost provable directly by
structural induction over the derivation. For example, consider the
case where the derivation ends in $\land R$\footnote{
  Traditionally one displays the analysed formula (in this case, $A
  \land B$) in the conclusion of the rule. This is occasionally a
  useful convention. We do not follow this convention here, instead,
  the requirement that the analysed formula appear in the conclusion
  is captured formally by a side condition (in this case, $A \land B
  \in \Delta$). Making the formula explicit leads to clumsy
  presentations of Craig's Interpolation Theorem, cf. Girard's
  presentation in \cite{girard87proof}. Incidentally, this
  presentation also witnesses our previous claim that informal proofs
  of Craig's Interpolation Theorem are prone to typos and other
  errors. This should not be understood as a failing on the part of
  Girard: he is one of the few who even attempt to detail the proof.

}.

\[
\begin{prooftree}
{\Gamma \vdash \Delta, A} \fskip {\Gamma \vdash \Delta, B}
\justifies
{\Gamma \vdash \Delta}
\using
{\land R}
\end{prooftree}
\]

We have that $A \land B \in \Delta$.
Using the induction hypothesis twice, we obtain a $C'$ such that

\[
\Gamma \vdash C' \skipand C' \vdash  \Delta, A
\]

and a $C''$ such that

\[
\Gamma \vdash C'' \skipand  C'' \vdash  \Delta, B
\]

Take $C = C' \land C''$. We have

\[
\begin{prooftree}
\[
\seq{\Gamma}{C'} 
\justifies
\seq{\Gamma}{C', C' \land C''}
\using
\WeakR
\]
\fskip
\[
\seq{\Gamma}{C''}
\justifies
\seq{\Gamma}{C'', C' \land C''}
\using
\WeakR
\]
\justifies
\seq{\Gamma}{C' \land C''}
\using
{\land R}
\end{prooftree}
\]

and

\[
\begin{prooftree}
\[
\[
\[
\seq{C'}{\Delta, A} 
\justifies
\seq{C'', C'}{\Delta,A}
\using
\WeakL
\]
\justifies
\seq{C' \land C'', C'', C'}{\Delta,A}
\using
\WeakL
\]
\justifies
\seq{C' \land C''}{\Delta,A}
\using
{\land L}
\]
\fskip
\[
\[
\[
\seq{C''}{\Delta, B} 
\justifies
\seq{C', C''}{\Delta,B}
\using
\WeakL
\]
\justifies
\seq{C' \land C'', C', C''}{\Delta,B}
\using
\WeakL
\]
\justifies
\seq{C' \land C''}{\Delta,B}
\using
{\land L}
\]
\justifies
\seq{C' \land C''}{\Delta}
\using
{\land R}
\end{prooftree}
\]

so that $\seq{\Gamma}{C}$ and $\seq{C}{\Delta}$.
Moreover, it is clear that the conditions on positive and negative
occurrences are satisfied.

For logical systems which do not include $\lnot, \limp$ connectives, the proof of Craig's Interpolation Theorem can be carried out straightforwardly.

However, for systems which include $\lnot, \limp$ the argument breaks down. The problem is that these connectives alter the polarity of the occurrences. For example, consider the case of $\lnot L$.

\[
\begin{prooftree}
{\Gamma \vdash \Delta, A}
\justifies
{\Gamma \vdash \Delta}
\using
{\lnot L}
\end{prooftree}
\]

We have $\lnot A \in \Gamma$. The induction hypothesis gives us a $C'$ such that 

\[
\Gamma \vdash C' \fskip \textit{and} \fskip C' \vdash \Delta, A
\]

and moreover satisfying the conditions on polarity of occurrences. However, one cannot directly obtain a $C$ such that 

\[
\Gamma \vdash C \fskip \textit{and} \fskip C \vdash \Delta
\]

because, for instance, $C'$ may contain a positive occurrence of $A$
(also occurring in $\Gamma$), whereas $A$ may not appear in $\Delta$,
so that $C$ may not contain a positive occurrence of $A$. 
Thus the direct approach to proving Craig's Interpolation Theorem
breaks down for polarity altering connectives.

The solution is to prove a stronger theorem. It is clear that the
problem lies with the polarity altering connectives such as $\lnot$.
It is reasonably easy to motivate a split sequent $\Gamma_1, \Gamma_2
\vdash \Delta_1, \Delta_2$. A goal sequent $\Gamma \vdash \Delta$ is
obtained by taking $\Gamma_1 = \Gamma, \Gamma_2 = \{\}, \Delta_1 =
\{\}, \Delta_2 = \Delta$. The additional components of the sequent,
$\Gamma_2, \Delta_1$ are used to keep track of the polarity changes
occurring in rules such as $\lnot L$.

\begin{theorem} (Strengthened Interpolation Theorem) \label{thmCraigStrong}
If 

\[
\seq{\Gamma_1,\Gamma_2}{\Delta_1,\Delta_2}
\]

then there exists a formula $C$ such that

\[
\seq{\Gamma_1}{\Delta_1,C} \skipand \seq{C,\Gamma_2}{\Delta_2}
\]

and moreover such that
\begin{itemize} 
\item Any predicate that occurs positively in $C$ occurs positively\footnote{``positively in $\Gamma_1, \lnot \Delta_1$'' means positively in $\Gamma_1$ or negatively in $\Delta_1$ etc.
} in $\Gamma_1, \lnot \Delta_1$ and positively in $\lnot \Gamma_2, \Delta_2$.
\item Any predicate that occurs negatively in $C$ occurs negatively in $\Gamma_1, \lnot \Delta_1$ and negatively in $\lnot \Gamma_2, \Delta_2$.
\end{itemize}
\end{theorem}

\isab{%
\endisatagproof
{\isafoldproof}%
\isadelimproof
\isanewline
\endisadelimproof
\isacommand{lemma}\isamarkupfalse%
\ craig{\isacharprime}{\isacharprime}{\isacharcolon}\ {\isachardoublequoteopen}\isanewline
\ \ {\isasymforall}\ d\ \ensuremath{{\isasymGamma}_{\isadigit{1}}}\ \ensuremath{{\isasymGamma}_{\isadigit{2}}}\ \ensuremath{{\isasymDelta}_{\isadigit{1}}}\ \ensuremath{{\isasymDelta}_{\isadigit{2}}}{\isachardot}\ \isanewline
\ \ \ \ is{\isacharunderscore}deriv\ d\ \isanewline
\ \ \ \ {\isasymand}\ root\ d\ {\isacharequal}\ {\isacharparenleft}\ensuremath{{\isasymGamma}_{\isadigit{1}}}\ {\isasymunion}\ \ensuremath{{\isasymGamma}_{\isadigit{2}}}{\isacharcomma}\ensuremath{{\isasymDelta}_{\isadigit{1}}}\ {\isasymunion}\ \ensuremath{{\isasymDelta}_{\isadigit{2}}}{\isacharparenright}\ \isanewline
\ \ \ \ {\isasymlongrightarrow}\ \isanewline
\ \ \ \ {\isacharparenleft}{\isasymexists}\ C{\isachardot}\isanewline
\ \ \ \ \ \ {\isacharparenleft}{\isasymexists}\ dl{\isachardot}\ is{\isacharunderscore}deriv\ dl\ {\isasymand}\ root\ dl\ {\isacharequal}\ {\isacharparenleft}\ensuremath{{\isasymGamma}_{\isadigit{1}}}{\isacharcomma}\ensuremath{{\isasymDelta}_{\isadigit{1}}}\ {\isasymunion}\ {\isacharbraceleft}C{\isacharbraceright}{\isacharparenright}{\isacharparenright}\isanewline
\ \ \ \ \ \ {\isasymand}\ {\isacharparenleft}{\isasymexists}\ dr{\isachardot}\ is{\isacharunderscore}deriv\ dr\ {\isasymand}\ root\ dr\ {\isacharequal}\ {\isacharparenleft}{\isacharbraceleft}C{\isacharbraceright}\ {\isasymunion}\ \ensuremath{{\isasymGamma}_{\isadigit{2}}}{\isacharcomma}\ensuremath{{\isasymDelta}_{\isadigit{2}}}{\isacharparenright}{\isacharparenright}\isanewline
\ \ \ \ \ \ {\isasymand}\ {\isacharparenleft}pos\ C\ {\isasymsubseteq}\ {\isacharparenleft}UNION\ \ensuremath{{\isasymGamma}_{\isadigit{1}}}\ pos{\isacharparenright}\ {\isasymunion}\ {\isacharparenleft}UNION\ \ensuremath{{\isasymDelta}_{\isadigit{1}}}\ neg{\isacharparenright}{\isacharparenright}\isanewline
\ \ \ \ \ \ {\isasymand}\ {\isacharparenleft}pos\ C\ {\isasymsubseteq}\ {\isacharparenleft}UNION\ \ensuremath{{\isasymGamma}_{\isadigit{2}}}\ neg{\isacharparenright}\ {\isasymunion}\ {\isacharparenleft}UNION\ \ensuremath{{\isasymDelta}_{\isadigit{2}}}\ pos{\isacharparenright}{\isacharparenright}\isanewline
\ \ \ \ \ \ {\isasymand}\ {\isacharparenleft}neg\ C\ {\isasymsubseteq}\ {\isacharparenleft}UNION\ \ensuremath{{\isasymGamma}_{\isadigit{1}}}\ neg{\isacharparenright}\ {\isasymunion}\ {\isacharparenleft}UNION\ \ensuremath{{\isasymDelta}_{\isadigit{1}}}\ pos{\isacharparenright}{\isacharparenright}\isanewline
\ \ \ \ \ \ {\isasymand}\ {\isacharparenleft}neg\ C\ {\isasymsubseteq}\ {\isacharparenleft}UNION\ \ensuremath{{\isasymGamma}_{\isadigit{2}}}\ pos{\isacharparenright}\ {\isasymunion}\ {\isacharparenleft}UNION\ \ensuremath{{\isasymDelta}_{\isadigit{2}}}\ neg{\isacharparenright}{\isacharparenright}{\isacharparenright}\isanewline
{\isachardoublequoteclose}\isanewline
}
The actual induction is a structural induction over the derivation of $\seq{\Gamma_1,\Gamma_2}{\Delta_1,\Delta_2}$. It is easiest to state this as an induction over the size of the derivation.

\isab{%
\endisatagproof
{\isafoldproof}%
\isadelimproof
\endisadelimproof
\isanewline
\isacommand{lemma}\isamarkupfalse%
\ craig{\isacharprime}{\isacharcolon}\ {\isachardoublequoteopen}\isanewline
\ \ {\isasymforall}\ n{\isachardot}\ {\isasymforall}\ d{\isachardot}\ size\ d\ {\isacharequal}\ n\ {\isasymlongrightarrow}\ \isanewline
\ \ \ \ {\isacharparenleft}{\isasymforall}\ \ensuremath{{\isasymGamma}_{\isadigit{1}}}\ \ensuremath{{\isasymGamma}_{\isadigit{2}}}\ \ensuremath{{\isasymDelta}_{\isadigit{1}}}\ \ensuremath{{\isasymDelta}_{\isadigit{2}}}{\isachardot}\ \isanewline
\ \ \ \ \ \ is{\isacharunderscore}deriv\ d\ \isanewline
\ \ \ \ \ \ {\isasymand}\ root\ d\ {\isacharequal}\ {\isacharparenleft}\ensuremath{{\isasymGamma}_{\isadigit{1}}}\ {\isasymunion}\ \ensuremath{{\isasymGamma}_{\isadigit{2}}}{\isacharcomma}\ensuremath{{\isasymDelta}_{\isadigit{1}}}\ {\isasymunion}\ \ensuremath{{\isasymDelta}_{\isadigit{2}}}{\isacharparenright}\ \isanewline
\ \ \ \ \ \ {\isasymlongrightarrow}\ \isanewline
\ \ \ \ \ \ {\isacharparenleft}{\isasymexists}\ C{\isachardot}\isanewline
\ \ \ \ \ \ \ \ {\isacharparenleft}{\isasymexists}\ dl{\isachardot}\ is{\isacharunderscore}deriv\ dl\ {\isasymand}\ root\ dl\ {\isacharequal}\ {\isacharparenleft}\ensuremath{{\isasymGamma}_{\isadigit{1}}}{\isacharcomma}\ensuremath{{\isasymDelta}_{\isadigit{1}}}\ {\isasymunion}\ {\isacharbraceleft}C{\isacharbraceright}{\isacharparenright}{\isacharparenright}\isanewline
\ \ \ \ \ \ \ \ {\isasymand}\ {\isacharparenleft}{\isasymexists}\ dr{\isachardot}\ is{\isacharunderscore}deriv\ dr\ {\isasymand}\ root\ dr\ {\isacharequal}\ {\isacharparenleft}{\isacharbraceleft}C{\isacharbraceright}\ {\isasymunion}\ \ensuremath{{\isasymGamma}_{\isadigit{2}}}{\isacharcomma}\ensuremath{{\isasymDelta}_{\isadigit{2}}}{\isacharparenright}{\isacharparenright}\isanewline
\ \ \ \ \ \ \ \ {\isasymand}\ {\isacharparenleft}pos\ C\ {\isasymsubseteq}\ {\isacharparenleft}UNION\ \ensuremath{{\isasymGamma}_{\isadigit{1}}}\ pos{\isacharparenright}\ {\isasymunion}\ {\isacharparenleft}UNION\ \ensuremath{{\isasymDelta}_{\isadigit{1}}}\ neg{\isacharparenright}{\isacharparenright}\isanewline
\ \ \ \ \ \ \ \ {\isasymand}\ {\isacharparenleft}pos\ C\ {\isasymsubseteq}\ {\isacharparenleft}UNION\ \ensuremath{{\isasymGamma}_{\isadigit{2}}}\ neg{\isacharparenright}\ {\isasymunion}\ {\isacharparenleft}UNION\ \ensuremath{{\isasymDelta}_{\isadigit{2}}}\ pos{\isacharparenright}{\isacharparenright}\isanewline
\ \ \ \ \ \ \ \ {\isasymand}\ {\isacharparenleft}neg\ C\ {\isasymsubseteq}\ {\isacharparenleft}UNION\ \ensuremath{{\isasymGamma}_{\isadigit{1}}}\ neg{\isacharparenright}\ {\isasymunion}\ {\isacharparenleft}UNION\ \ensuremath{{\isasymDelta}_{\isadigit{1}}}\ pos{\isacharparenright}{\isacharparenright}\isanewline
\ \ \ \ \ \ \ \ {\isasymand}\ {\isacharparenleft}neg\ C\ {\isasymsubseteq}\ {\isacharparenleft}UNION\ \ensuremath{{\isasymGamma}_{\isadigit{2}}}\ pos{\isacharparenright}\ {\isasymunion}\ {\isacharparenleft}UNION\ \ensuremath{{\isasymDelta}_{\isadigit{2}}}\ neg{\isacharparenright}{\isacharparenright}{\isacharparenright}{\isacharparenright}\isanewline
{\isachardoublequoteclose}\isanewline
}
\begin{corollary} (Craig's Interpolation Theorem)
\end{corollary}
\begin{proof} The original formulation of Craig's Interpolation Theorem follows immediately from Thm. \ref{thmCraigStrong} by taking 
$(\Gamma,\{\},\{\},\Delta)=(\Gamma_1,\Gamma_2,\Delta_1,\Delta_2)$.
\end{proof}

%%%%%%%%%%%%%%%%%%%%%%%%%%%%%%%%%%%%%%%%%%%%%%%%%%%%%%%%%%%%%%%%%%%%%%%%%%%%%%%%

\tsection{Proof of Craig's Interpolation Theorem}

\label{sectCraigProof}

We aim to prove the strengthened form of Craig's Interpolation
Theorem, Thm. \ref{thmCraigStrong}. We induct over the size of the
derivation $d$ of $\seq{\Gamma_1,\Gamma_2}{\Delta_1,\Delta_2}$, so
that we can use the induction hypothesis for all derivations of
smaller size. The body of the proof proceeds by a case analysis on the
last constructor of the given derivation. 

In the following cases, apart from the {\Init} case, we do not check
the conditions regarding positive and negative occurrences in the
interpolation formula. These conditions are straightforward to verify.

%%%%%%%%%%%%%%%%%%%%%%%%%%%%%%%%%%%%%%%%
\tsubsection{Case {\Init}}

We give a formal Isar rendition of the case $d$ ends in rule {\Init}.
There are four subcases. We give a full rendition of the first
subcase. The 3 remaining subcases are very similar. We provide the
explicit $C$ for these cases but suppress the mundane proofs.  The
full details of the remaining subcases can be found in the mechanised theory
script.

\isab{%
\endisatagproof
{\isafoldproof}%
\isadelimproof
\isanewline
\endisadelimproof
\isacommand{lemma}\isamarkupfalse%
\ \isakeyword{assumes}\ a{\isacharcolon}\ {\isachardoublequoteopen}is{\isacharunderscore}deriv\ d{\isachardoublequoteclose}\ \isakeyword{and}\ b{\isacharcolon}\ {\isachardoublequoteopen}root\ d\ {\isacharequal}\ {\isacharparenleft}\ensuremath{{\isasymGamma}_{\isadigit{1}}}\ {\isasymunion}\ \ensuremath{{\isasymGamma}_{\isadigit{2}}}{\isacharcomma}\ \ensuremath{{\isasymDelta}_{\isadigit{1}}}\ {\isasymunion}\ \ensuremath{{\isasymDelta}_{\isadigit{2}}}{\isacharparenright}{\isachardoublequoteclose}\ \isakeyword{and}\ c{\isacharcolon}\ {\isachardoublequoteopen}d\ {\isacharequal}\ Init\ {\isasymGamma}{\isasymDelta}{\isachardoublequoteclose}\isanewline
\ \ \isakeyword{shows}\ {\isachardoublequoteopen}{\isasymexists}\ C{\isachardot}\ {\isacharparenleft}{\isasymexists}\ dl{\isachardot}\ is{\isacharunderscore}deriv\ dl\ {\isasymand}\ root\ dl\ {\isacharequal}\ {\isacharparenleft}\ensuremath{{\isasymGamma}_{\isadigit{1}}}{\isacharcomma}\ \ensuremath{{\isasymDelta}_{\isadigit{1}}}\ {\isasymunion}\ {\isacharbraceleft}C{\isacharbraceright}{\isacharparenright}{\isacharparenright}\ \isanewline
\ \ \ \ {\isasymand}\ {\isacharparenleft}{\isasymexists}\ dr{\isachardot}\ is{\isacharunderscore}deriv\ dr\ {\isasymand}\ root\ dr\ {\isacharequal}\ {\isacharparenleft}{\isacharbraceleft}C{\isacharbraceright}\ {\isasymunion}\ \ensuremath{{\isasymGamma}_{\isadigit{2}}}{\isacharcomma}\ \ensuremath{{\isasymDelta}_{\isadigit{2}}}{\isacharparenright}{\isacharparenright}\ \isanewline
\ \ \ \ {\isasymand}\ {\isacharparenleft}pos\ C\ {\isasymsubseteq}\ {\isacharparenleft}UNION\ \ensuremath{{\isasymGamma}_{\isadigit{1}}}\ pos{\isacharparenright}\ {\isasymunion}\ {\isacharparenleft}UNION\ \ensuremath{{\isasymDelta}_{\isadigit{1}}}\ neg{\isacharparenright}{\isacharparenright}\isanewline
\ \ \ \ {\isasymand}\ {\isacharparenleft}pos\ C\ {\isasymsubseteq}\ {\isacharparenleft}UNION\ \ensuremath{{\isasymGamma}_{\isadigit{2}}}\ neg{\isacharparenright}\ {\isasymunion}\ {\isacharparenleft}UNION\ \ensuremath{{\isasymDelta}_{\isadigit{2}}}\ pos{\isacharparenright}{\isacharparenright}\isanewline
\ \ \ \ {\isasymand}\ {\isacharparenleft}neg\ C\ {\isasymsubseteq}\ {\isacharparenleft}UNION\ \ensuremath{{\isasymGamma}_{\isadigit{1}}}\ neg{\isacharparenright}\ {\isasymunion}\ {\isacharparenleft}UNION\ \ensuremath{{\isasymDelta}_{\isadigit{1}}}\ pos{\isacharparenright}{\isacharparenright}\isanewline
\ \ \ \ {\isasymand}\ {\isacharparenleft}neg\ C\ {\isasymsubseteq}\ {\isacharparenleft}UNION\ \ensuremath{{\isasymGamma}_{\isadigit{2}}}\ pos{\isacharparenright}\ {\isasymunion}\ {\isacharparenleft}UNION\ \ensuremath{{\isasymDelta}_{\isadigit{2}}}\ neg{\isacharparenright}{\isacharparenright}{\isachardoublequoteclose}\isanewline
\ \ {\isacharparenleft}\isakeyword{is}\ {\isachardoublequoteopen}{\isasymexists}\ C{\isachardot}\ {\isacharquery}P\ C{\isachardoublequoteclose}\ {\isacharparenright}\isanewline
\isadelimproof
\endisadelimproof
\isatagproof
\isacommand{proof}\isamarkupfalse%
\ {\isacharminus}\isanewline
\ \ \isacommand{from}\isamarkupfalse%
\ a\ b\ c\ \isacommand{obtain}\isamarkupfalse%
\ A\ \isakeyword{where}\ {\isachardoublequoteopen}{\isacharparenleft}A\ {\isasymin}\ \ensuremath{{\isasymGamma}_{\isadigit{1}}}\ {\isasymand}\ A\ {\isasymin}\ \ensuremath{{\isasymDelta}_{\isadigit{1}}}{\isacharparenright}\ {\isasymor}\ {\isacharparenleft}A\ {\isasymin}\ \ensuremath{{\isasymGamma}_{\isadigit{1}}}\ {\isasymand}\ A\ {\isasymin}\ \ensuremath{{\isasymDelta}_{\isadigit{2}}}{\isacharparenright}\ {\isasymor}\ {\isacharparenleft}A\ {\isasymin}\ \ensuremath{{\isasymGamma}_{\isadigit{2}}}\ {\isasymand}\ A\ {\isasymin}\ \ensuremath{{\isasymDelta}_{\isadigit{1}}}{\isacharparenright}\ {\isasymor}\ {\isacharparenleft}A\ {\isasymin}\ \ensuremath{{\isasymGamma}_{\isadigit{2}}}\ {\isasymand}\ A\ {\isasymin}\ \ensuremath{{\isasymDelta}_{\isadigit{2}}}{\isacharparenright}{\isachardoublequoteclose}\ \isanewline
\ \ \isacommand{thus}\isamarkupfalse%
\ {\isacharquery}thesis\isanewline
\ \ \isacommand{proof}\isamarkupfalse%
\ {\isacharparenleft}elim\ disjE{\isacharparenright}\isanewline
\ \ \ \ \isacommand{assume}\isamarkupfalse%
\ {\isachardoublequoteopen}A\ {\isasymin}\ \ensuremath{{\isasymGamma}_{\isadigit{1}}}\ {\isasymand}\ A\ {\isasymin}\ \ensuremath{{\isasymDelta}_{\isadigit{1}}}{\isachardoublequoteclose}\isanewline
\ \ \ \ \isacommand{have}\isamarkupfalse%
\ {\isachardoublequoteopen}{\isacharquery}P\ \ensuremath{\underline{\isasymbottom}}{\isachardoublequoteclose}\isanewline
\ \ \ \ \isacommand{proof}\isamarkupfalse%
\ {\isacharparenleft}intro\ conjI{\isacharparenright}\isanewline
\ \ \ \ \ \ \isacommand{show}\isamarkupfalse%
\ {\isachardoublequoteopen}{\isacharparenleft}{\isasymexists}\ dl{\isachardot}\ is{\isacharunderscore}deriv\ dl\ {\isasymand}\ root\ dl\ {\isacharequal}\ {\isacharparenleft}\ensuremath{{\isasymGamma}_{\isadigit{1}}}{\isacharcomma}\ \ensuremath{{\isasymDelta}_{\isadigit{1}}}\ {\isasymunion}\ {\isacharbraceleft}\ensuremath{\underline{\isasymbottom}}{\isacharbraceright}{\isacharparenright}{\isacharparenright}{\isachardoublequoteclose}\isanewline
\ \ \ \ \ \ \isacommand{proof}\isamarkupfalse%
\ \isanewline
\ \ \ \ \ \ \ \ \isacommand{let}\isamarkupfalse%
\ {\isacharquery}dl\ {\isacharequal}\ {\isachardoublequoteopen}Init\ {\isacharparenleft}\ensuremath{{\isasymGamma}_{\isadigit{1}}}{\isacharcomma}\ensuremath{{\isasymDelta}_{\isadigit{1}}}\ {\isasymunion}\ {\isacharbraceleft}\ensuremath{\underline{\isasymbottom}}{\isacharbraceright}{\isacharparenright}{\isachardoublequoteclose}\isanewline
\ \ \ \ \ \ \ \ \isacommand{show}\isamarkupfalse%
\ {\isachardoublequoteopen}is{\isacharunderscore}deriv\ {\isacharquery}dl\ {\isasymand}\ root\ {\isacharquery}dl\ {\isacharequal}\ {\isacharparenleft}\ensuremath{{\isasymGamma}_{\isadigit{1}}}{\isacharcomma}\ \ensuremath{{\isasymDelta}_{\isadigit{1}}}\ {\isasymunion}\ {\isacharbraceleft}\ensuremath{\underline{\isasymbottom}}{\isacharbraceright}{\isacharparenright}{\isachardoublequoteclose}\ \isacommand{by}\isamarkupfalse%
{\isacharparenleft}force{\isacharbang}\ simp\ add{\isacharcolon}\ Let{\isacharunderscore}def{\isacharparenright}\isanewline
\ \ \ \ \ \ \isacommand{qed}\isamarkupfalse%
\isanewline
\ \ \ \ \isacommand{next}\isamarkupfalse%
\isanewline
\ \ \ \ \ \ \isacommand{show}\isamarkupfalse%
\ {\isachardoublequoteopen}{\isacharparenleft}{\isasymexists}\ dr{\isachardot}\ is{\isacharunderscore}deriv\ dr\ {\isasymand}\ root\ dr\ {\isacharequal}\ {\isacharparenleft}{\isacharbraceleft}\ensuremath{\underline{\isasymbottom}}{\isacharbraceright}\ {\isasymunion}\ \ensuremath{{\isasymGamma}_{\isadigit{2}}}{\isacharcomma}\ \ensuremath{{\isasymDelta}_{\isadigit{2}}}{\isacharparenright}{\isacharparenright}{\isachardoublequoteclose}\isanewline
\ \ \ \ \ \ \isacommand{proof}\isamarkupfalse%
\isanewline
\ \ \ \ \ \ \ \ \isacommand{let}\isamarkupfalse%
\ {\isacharquery}dr\ {\isacharequal}\ {\isachardoublequoteopen}\ensuremath{\underline{\isasymbottom}}L\ {\isacharparenleft}{\isacharbraceleft}\ensuremath{\underline{\isasymbottom}}{\isacharbraceright}\ {\isasymunion}\ \ensuremath{{\isasymGamma}_{\isadigit{2}}}{\isacharcomma}\ensuremath{{\isasymDelta}_{\isadigit{2}}}{\isacharparenright}{\isachardoublequoteclose}\isanewline
\ \ \ \ \ \ \ \ \isacommand{show}\isamarkupfalse%
\ {\isachardoublequoteopen}is{\isacharunderscore}deriv\ {\isacharquery}dr\ {\isasymand}\ root\ {\isacharquery}dr\ {\isacharequal}\ {\isacharparenleft}{\isacharbraceleft}\ensuremath{\underline{\isasymbottom}}{\isacharbraceright}\ {\isasymunion}\ \ensuremath{{\isasymGamma}_{\isadigit{2}}}{\isacharcomma}\ \ensuremath{{\isasymDelta}_{\isadigit{2}}}{\isacharparenright}{\isachardoublequoteclose}\ \isacommand{by}\isamarkupfalse%
{\isacharparenleft}force{\isacharbang}\ simp\ add{\isacharcolon}\ Let{\isacharunderscore}def{\isacharparenright}\isanewline
\ \ \ \ \ \ \isacommand{qed}\isamarkupfalse%
\isanewline
\ \ \ \ \isacommand{next}\isamarkupfalse%
\isanewline
\ \ \ \ \ \ \isacommand{show}\isamarkupfalse%
\ {\isachardoublequoteopen}pos\ \ensuremath{\underline{\isasymbottom}}\ {\isasymsubseteq}\ {\isacharparenleft}UNION\ \ensuremath{{\isasymGamma}_{\isadigit{1}}}\ pos{\isacharparenright}\ {\isasymunion}\ {\isacharparenleft}UNION\ \ensuremath{{\isasymDelta}_{\isadigit{1}}}\ neg{\isacharparenright}{\isachardoublequoteclose}\ \isacommand{by}\isamarkupfalse%
{\isacharparenleft}simp{\isacharbang}{\isacharparenright}\isanewline
\ \ \ \ \ \ \isacommand{show}\isamarkupfalse%
\ {\isachardoublequoteopen}pos\ \ensuremath{\underline{\isasymbottom}}\ {\isasymsubseteq}\ {\isacharparenleft}UNION\ \ensuremath{{\isasymGamma}_{\isadigit{2}}}\ neg{\isacharparenright}\ {\isasymunion}\ {\isacharparenleft}UNION\ \ensuremath{{\isasymDelta}_{\isadigit{2}}}\ pos{\isacharparenright}{\isachardoublequoteclose}\ \isacommand{by}\isamarkupfalse%
{\isacharparenleft}simp{\isacharbang}{\isacharparenright}\isanewline
\ \ \ \ \ \ \isacommand{show}\isamarkupfalse%
\ {\isachardoublequoteopen}neg\ \ensuremath{\underline{\isasymbottom}}\ {\isasymsubseteq}\ {\isacharparenleft}UNION\ \ensuremath{{\isasymGamma}_{\isadigit{1}}}\ neg{\isacharparenright}\ {\isasymunion}\ {\isacharparenleft}UNION\ \ensuremath{{\isasymDelta}_{\isadigit{1}}}\ pos{\isacharparenright}{\isachardoublequoteclose}\ \isacommand{by}\isamarkupfalse%
{\isacharparenleft}simp{\isacharbang}{\isacharparenright}\isanewline
\ \ \ \ \ \ \isacommand{show}\isamarkupfalse%
\ {\isachardoublequoteopen}neg\ \ensuremath{\underline{\isasymbottom}}\ {\isasymsubseteq}\ {\isacharparenleft}UNION\ \ensuremath{{\isasymGamma}_{\isadigit{2}}}\ pos{\isacharparenright}\ {\isasymunion}\ {\isacharparenleft}UNION\ \ensuremath{{\isasymDelta}_{\isadigit{2}}}\ neg{\isacharparenright}{\isachardoublequoteclose}\ \isacommand{by}\isamarkupfalse%
{\isacharparenleft}simp{\isacharbang}{\isacharparenright}\isanewline
\ \ \ \ \isacommand{qed}\isamarkupfalse%
\isanewline
\ \ \ \ \isacommand{thus}\isamarkupfalse%
\ {\isacharquery}thesis\ \isacommand{{\isachardot}{\isachardot}}\isamarkupfalse%
\isanewline
\ \ \isacommand{next}\isamarkupfalse%
\isanewline
\ \ \ \ \isacommand{assume}\isamarkupfalse%
\ {\isachardoublequoteopen}A\ {\isasymin}\ \ensuremath{{\isasymGamma}_{\isadigit{1}}}\ {\isasymand}\ A\ {\isasymin}\ \ensuremath{{\isasymDelta}_{\isadigit{2}}}{\isachardoublequoteclose}\isanewline
\ \ \ \ \isacommand{have}\isamarkupfalse%
\ {\isachardoublequoteopen}{\isacharquery}P\ A{\isachardoublequoteclose}\ \isanewline
\ \ \ \ \isacommand{thus}\isamarkupfalse%
\ {\isacharquery}thesis\ \isacommand{{\isachardot}{\isachardot}}\isamarkupfalse%
\isanewline
\ \ \isacommand{next}\isamarkupfalse%
\isanewline
\ \ \ \ \isacommand{assume}\isamarkupfalse%
\ {\isachardoublequoteopen}A\ {\isasymin}\ \ensuremath{{\isasymGamma}_{\isadigit{2}}}\ {\isasymand}\ A\ {\isasymin}\ \ensuremath{{\isasymDelta}_{\isadigit{1}}}{\isachardoublequoteclose}\isanewline
\ \ \ \ \isacommand{have}\isamarkupfalse%
\ {\isachardoublequoteopen}{\isacharquery}P\ {\isacharparenleft}\ensuremath{\underline{\isasymnot}}\ A{\isacharparenright}{\isachardoublequoteclose}\ \isanewline
\ \ \ \ \isacommand{thus}\isamarkupfalse%
\ {\isacharquery}thesis\ \isacommand{{\isachardot}{\isachardot}}\isamarkupfalse%
\isanewline
\ \ \isacommand{next}\isamarkupfalse%
\isanewline
\ \ \ \ \isacommand{assume}\isamarkupfalse%
\ {\isachardoublequoteopen}A\ {\isasymin}\ \ensuremath{{\isasymGamma}_{\isadigit{2}}}\ {\isasymand}\ A\ {\isasymin}\ \ensuremath{{\isasymDelta}_{\isadigit{2}}}{\isachardoublequoteclose}\isanewline
\ \ \ \ \isacommand{have}\isamarkupfalse%
\ {\isachardoublequoteopen}{\isacharquery}P\ \ensuremath{\underline{\isasymtop}}{\isachardoublequoteclose}\ \isanewline
\ \ \ \ \isacommand{thus}\isamarkupfalse%
\ {\isacharquery}thesis\ \isacommand{{\isachardot}{\isachardot}}\isamarkupfalse%
\isanewline
\ \ \isacommand{qed}\isamarkupfalse%
\isanewline
\isacommand{qed}\isamarkupfalse%
\isanewline
}
%%%%%%%%%%%%%%%%%%%%%%%%%%%%%%%%%%%%%%%%%%%%%%%%%%%%%%%%%%%%%%%%%%%%%%%%%%%%%%%%
\tsubsection{Case $\land L$}

We have 

\newcommand{\newover}[2]{\[ #1 \thickness=0.0em\justifies #2\]}

\[
\begin{prooftree}
\newover{d}{\seq{A,B,\Gamma_1,\Gamma_2}{\Delta_1,\Delta_2}}
\justifies
\seq{\Gamma_1,\Gamma_2}{\Delta_1,\Delta_2}
\using
{\land L}
\end{prooftree}
\]

From wellformedness of the derivation, we have that $A \land B \in
\Gamma_1,\Gamma_2$. There are two subcases, $A \land B \in \Gamma_1$
or $A \land B \in \Gamma_2$.

\begin{itemize}
%%%%%%%%%%%%%%%%%%%%%%%%%%%%%%%%%%%%%%%%
\item Case $A \land B \in \Gamma_1$. The I.H. applied to $d$ gives $C', dl, dr$ such that 

\[
\begin{prooftree}
\newover{dl}{\seq{A,B,\Gamma_1}{\Delta_1,C'}}
\skipand 
\newover{dr}{\seq{C', \Gamma_2}{\Delta_2}}
\thickness0em
\justifies
\end{prooftree}
\]

Take $C = C'$. Then 

\[
\begin{prooftree}
\newover{dl}{\seq{A,B,\Gamma_1}{\Delta_1,C'}}
\justifies
\seq{\Gamma_1}{\Delta_1,C'}
\using
{\land L}
\end{prooftree}
\]

The formal proof witness is 

\isab{\begin{isabelle}%
{\isachardoublequote}\isanewline
\ensuremath{\underline{\isasymand}}L\ {\isacharparenleft}\ensuremath{{\isasymGamma}_{\isadigit{1}}}{\isacharcomma}\ensuremath{{\isasymDelta}_{\isadigit{1}}}\ {\isasymunion}\ {\isacharbraceleft}C{\isacharprime}{\isacharbraceright}{\isacharparenright}\ dl\isanewline
{\isachardoublequote}%
\end{isabelle}
}
$dr$ is already a witness for $\seq{C', \Gamma_2}{\Delta_2}$.

%%%%%%%%%%%%%%%%%%%%%%%%%%%%%%%%%%%%%%%%
\item Case $A \land B \in \Gamma_2$. The I.H. applied to $d$ gives $C', dl, dr$ such that 

\[
\begin{prooftree}
\newover{dl}{\seq{\Gamma_1}{\Delta_1,C'}}
\skipand 
\newover{dr}{\seq{C',A,B, \Gamma_2}{\Delta_2}}
\thickness0em
\justifies
\end{prooftree}
\]

Take $C = C'$. Then 

\[
\begin{prooftree}
\newover{dr}{\seq{C',A,B,\Gamma_2}{\Delta_2}}
\justifies
\seq{C',\Gamma_2}{\Delta_2}
\using
{\land L}
\end{prooftree}
\]

The formal proof witness is 

\isab{\begin{isabelle}%
{\isachardoublequote}\isanewline
\ensuremath{\underline{\isasymand}}L\ {\isacharparenleft}{\isacharbraceleft}C{\isacharprime}{\isacharbraceright}\ {\isasymunion}\ \ensuremath{{\isasymGamma}_{\isadigit{2}}}{\isacharcomma}\ensuremath{{\isasymDelta}_{\isadigit{2}}}{\isacharparenright}\ dr\isanewline
{\isachardoublequote}%
\end{isabelle}
}
$dl$ is already a witness for $\seq{\Gamma_1}{\Delta_1,C'}$.

\end{itemize}

%%%%%%%%%%%%%%%%%%%%%%%%%%%%%%%%%%%%%%%%%%%%%%%%%%%%%%%%%%%%%%%%%%%%%%%%%%%%%%%%
\tsubsection{Case $\land R$}

We have 

\[
\begin{prooftree}
\newover{dl}{\seq{\Gamma_1,\Gamma_2}{\Delta_1,\Delta_2,A}}
\fskip
\newover{dr}{\seq{\Gamma_1,\Gamma_2}{\Delta_1,\Delta_2,B}}
\justifies
\seq{\Gamma_1,\Gamma_2}{\Delta_1,\Delta_2}
\using
{\land R}
\end{prooftree}
\]

From wellformedness of the derivation, we have that $A \land B \in
\Delta_1,\Delta_2$. There are two subcases, $A \land B \in \Delta_1$
or $A \land B \in \Delta_2$.

\begin{itemize}
%%%%%%%%%%%%%%%%%%%%%%%%%%%%%%%%%%%%%%%%
\item Case $A \land B \in \Delta_1$. The I.H. applied to $dl$ gives $C',dll,dlr$ such that

\[
\begin{prooftree}
\newover{dll}{\seq{\Gamma_1}{\Delta_1,A,C'}}
\skipand
\newover{dlr}{\seq{C',\Gamma_2}{\Delta_2}}
\thickness0em
\justifies
\end{prooftree}
\]

The I.H. applied to $dr$ gives $C'',drl,drr$ such that 

\[
\begin{prooftree}
\newover{drl}{\seq{\Gamma_1}{\Delta_1,B,C''}}
\skipand
\newover{drr}{\seq{C'',\Gamma_2}{\Delta_2}}
\thickness0em
\justifies
\end{prooftree}
\]

Take $C = C' \lor C''$. Then 

\begin{scriptsize}
\[
\begin{prooftree}
\[
\[
\[
\newover{dll}{\seq{\Gamma_1}{\Delta_1,A,C'}}
\justifies
\seq{\Gamma_1}{\Delta_1,A,C',C' \lor C''}
\using
\WeakR
\]
\justifies
\seq{\Gamma_1}{\Delta_1,A,C',C' \lor C'',C''}
\using
\WeakR
\]
\justifies
\seq{\Gamma_1}{\Delta_1,A, C' \lor C''}
\using
{\lor R}
\]
\[
\[
\[
\newover{drl}{\seq{\Gamma_1}{\Delta_1,B,C''}}
\justifies
\seq{\Gamma_1}{\Delta_1,B,C'',C' \lor C''}
\using
\WeakR
\]
\justifies
\seq{\Gamma_1}{\Delta_1,B,C'',C' \lor C'',C'}
\using
\WeakR
\]
\justifies
\seq{\Gamma_1}{\Delta_1,B, C' \lor C''}
\using
{\lor R}
\]
\justifies
\seq{\Gamma_1}{\Delta_1, C' \lor C''}
\using
{\land R}
\end{prooftree}
\]
\end{scriptsize}

The formal proof witness is

\isab{\begin{isabelle}%
{\isachardoublequote}\isanewline
\ \ \ \ let\ dll{\isacharprime}\ {\isacharequal}\ WR\ {\isacharparenleft}\ensuremath{{\isasymGamma}_{\isadigit{1}}}{\isacharcomma}\ \ensuremath{{\isasymDelta}_{\isadigit{1}}}\ {\isasymunion}\ {\isacharbraceleft}A{\isacharcomma}\ C{\isacharprime}{\isacharcomma}\ \ensuremath{\underline{\isasymor}}\ C{\isacharprime}\ C{\isacharprime}{\isacharprime}{\isacharbraceright}{\isacharparenright}\ dll\ in\isanewline
\ \ \ \ let\ dll{\isacharprime}{\isacharprime}\ {\isacharequal}\ WR\ {\isacharparenleft}\ensuremath{{\isasymGamma}_{\isadigit{1}}}{\isacharcomma}\ \ensuremath{{\isasymDelta}_{\isadigit{1}}}\ {\isasymunion}\ {\isacharbraceleft}A{\isacharcomma}\ C{\isacharprime}{\isacharcomma}\ \ensuremath{\underline{\isasymor}}\ C{\isacharprime}\ C{\isacharprime}{\isacharprime}{\isacharcomma}\ C{\isacharprime}{\isacharprime}{\isacharbraceright}{\isacharparenright}\ dll{\isacharprime}\ in\isanewline
\ \ \ \ let\ dll{\isacharprime}{\isacharprime}{\isacharprime}\ {\isacharequal}\ \ensuremath{\underline{\isasymor}}R\ {\isacharparenleft}\ensuremath{{\isasymGamma}_{\isadigit{1}}}{\isacharcomma}\ \ensuremath{{\isasymDelta}_{\isadigit{1}}}\ {\isasymunion}\ {\isacharbraceleft}A{\isacharcomma}\ensuremath{\underline{\isasymor}}\ C{\isacharprime}\ C{\isacharprime}{\isacharprime}{\isacharbraceright}{\isacharparenright}\ dll{\isacharprime}{\isacharprime}\ in\isanewline
\ \ \ \ \isanewline
\ \ \ \ let\ drl{\isacharprime}\ {\isacharequal}\ WR\ {\isacharparenleft}\ensuremath{{\isasymGamma}_{\isadigit{1}}}{\isacharcomma}\ \ensuremath{{\isasymDelta}_{\isadigit{1}}}\ {\isasymunion}\ {\isacharbraceleft}B{\isacharcomma}\ C{\isacharprime}{\isacharprime}{\isacharcomma}\ \ensuremath{\underline{\isasymor}}\ C{\isacharprime}\ C{\isacharprime}{\isacharprime}{\isacharbraceright}{\isacharparenright}\ drl\ in\isanewline
\ \ \ \ let\ drl{\isacharprime}{\isacharprime}\ {\isacharequal}\ WR\ {\isacharparenleft}\ensuremath{{\isasymGamma}_{\isadigit{1}}}{\isacharcomma}\ \ensuremath{{\isasymDelta}_{\isadigit{1}}}\ {\isasymunion}\ {\isacharbraceleft}B{\isacharcomma}\ C{\isacharprime}{\isacharprime}{\isacharcomma}\ \ensuremath{\underline{\isasymor}}\ C{\isacharprime}\ C{\isacharprime}{\isacharprime}{\isacharcomma}\ C{\isacharprime}{\isacharbraceright}{\isacharparenright}\ drl{\isacharprime}\ in\isanewline
\ \ \ \ let\ drl{\isacharprime}{\isacharprime}{\isacharprime}\ {\isacharequal}\ \ensuremath{\underline{\isasymor}}R\ {\isacharparenleft}\ensuremath{{\isasymGamma}_{\isadigit{1}}}{\isacharcomma}\ \ensuremath{{\isasymDelta}_{\isadigit{1}}}\ {\isasymunion}\ {\isacharbraceleft}B{\isacharcomma}\ensuremath{\underline{\isasymor}}\ C{\isacharprime}\ C{\isacharprime}{\isacharprime}{\isacharbraceright}{\isacharparenright}\ drl{\isacharprime}{\isacharprime}\ in\isanewline
\ \ \ \ \ \ \ensuremath{\underline{\isasymand}}R\ {\isacharparenleft}\ensuremath{{\isasymGamma}_{\isadigit{1}}}{\isacharcomma}\ \ensuremath{{\isasymDelta}_{\isadigit{1}}}\ {\isasymunion}\ {\isacharbraceleft}\ensuremath{\underline{\isasymor}}\ C{\isacharprime}\ C{\isacharprime}{\isacharprime}{\isacharbraceright}{\isacharparenright}\ dll{\isacharprime}{\isacharprime}{\isacharprime}\ drl{\isacharprime}{\isacharprime}{\isacharprime}\isanewline
{\isachardoublequote}%
\end{isabelle}
}
Similarly

\[
\begin{prooftree}
\[
\newover{dlr}{\seq{C',\Gamma_2}{\Delta_2}}
\justifies
\seq{C' \lor C'', C', \Gamma_2}{\Delta_2}
\using
\WeakL
\]
\[
\newover{drr}{\seq{C'',\Gamma_2}{\Delta_2}}
\justifies
\seq{C' \lor C'', C'',\Gamma_2}{\Delta_2}
\using
\WeakL
\]
\justifies
\seq{C' \lor C'', \Gamma_2}{\Delta_2}
\using
{\lor L}
\end{prooftree}
\]

The formal proof witness is

\isab{\begin{isabelle}%
{\isachardoublequote}\isanewline
\ \ \ \ let\ dlr{\isacharprime}\ {\isacharequal}\ WL\ {\isacharparenleft}{\isacharbraceleft}\ensuremath{\underline{\isasymor}}\ C{\isacharprime}\ C{\isacharprime}{\isacharprime}{\isacharcomma}\ C{\isacharprime}{\isacharbraceright}\ {\isasymunion}\ \ensuremath{{\isasymGamma}_{\isadigit{2}}}{\isacharcomma}\ \ensuremath{{\isasymDelta}_{\isadigit{2}}}{\isacharparenright}\ dlr\ in\isanewline
\ \ \ \ let\ drr{\isacharprime}\ {\isacharequal}\ WL\ {\isacharparenleft}{\isacharbraceleft}\ensuremath{\underline{\isasymor}}\ C{\isacharprime}\ C{\isacharprime}{\isacharprime}{\isacharcomma}\ C{\isacharprime}{\isacharprime}{\isacharbraceright}\ {\isasymunion}\ \ensuremath{{\isasymGamma}_{\isadigit{2}}}{\isacharcomma}\ \ensuremath{{\isasymDelta}_{\isadigit{2}}}{\isacharparenright}\ drr\ in\isanewline
\ \ \ \ \ \ \ensuremath{\underline{\isasymor}}L\ {\isacharparenleft}{\isacharbraceleft}\ensuremath{\underline{\isasymor}}\ C{\isacharprime}\ C{\isacharprime}{\isacharprime}{\isacharbraceright}\ {\isasymunion}\ \ensuremath{{\isasymGamma}_{\isadigit{2}}}{\isacharcomma}\ \ensuremath{{\isasymDelta}_{\isadigit{2}}}{\isacharparenright}\ dlr{\isacharprime}\ drr{\isacharprime}\isanewline
{\isachardoublequote}%
\end{isabelle}
}
%%%%%%%%%%%%%%%%%%%%%%%%%%%%%%%%%%%%%%%%
\item Case $A \land B \in \Delta_2$. The I.H. applied to $dl$ gives $C',dll,dlr$ such that 

\[
\begin{prooftree}
\newover{dll}{\seq{\Gamma_1}{\Delta_1,C'}}
\skipand
\newover{dlr}{\seq{C',\Gamma_2}{\Delta_2,A}}
\thickness0em
\justifies
\end{prooftree}
\]

The I.H. applied to $dr$ gives $C'',drl,drr$ such that 

\[
\begin{prooftree}
\newover{drl}{\seq{\Gamma_1}{\Delta_1,C''}}
\skipand
\newover{drr}{\seq{C'',\Gamma_2}{\Delta_2,B}}
\thickness0em
\justifies
\end{prooftree}
\]

Take $C = C' \land C''$. Then 

\[
\begin{prooftree}
\[
\newover{dll}{\seq{\Gamma_1}{\Delta_1,C'}}
\justifies
\seq{\Gamma_1}{\Delta_1,C',C' \land C''}
\using
\WeakR
\]
\[
\newover{drl}{\seq{\Gamma_1}{\Delta_1,C''}}
\justifies
{\seq{\Gamma_1}{\Delta_1,C'',C' \land C''}}
\using
\WeakR
\]
\justifies
\seq{\Gamma_1}{\Delta_1, C' \land C''}
\using
{\land R}
\end{prooftree}
\]

The formal proof witness is

\isab{\begin{isabelle}%
{\isachardoublequote}\isanewline
\ \ \ \ let\ dll{\isacharprime}\ {\isacharequal}\ WR\ {\isacharparenleft}\ensuremath{{\isasymGamma}_{\isadigit{1}}}{\isacharcomma}\ensuremath{{\isasymDelta}_{\isadigit{1}}}\ {\isasymunion}\ {\isacharbraceleft}C{\isacharprime}{\isacharcomma}\ensuremath{\underline{\isasymand}}\ C{\isacharprime}\ C{\isacharprime}{\isacharprime}{\isacharbraceright}{\isacharparenright}\ dll\ in\ \isanewline
\ \ \ \ let\ drl{\isacharprime}\ {\isacharequal}\ WR\ {\isacharparenleft}\ensuremath{{\isasymGamma}_{\isadigit{1}}}{\isacharcomma}\ensuremath{{\isasymDelta}_{\isadigit{1}}}\ {\isasymunion}\ {\isacharbraceleft}C{\isacharprime}{\isacharprime}{\isacharcomma}\ensuremath{\underline{\isasymand}}\ C{\isacharprime}\ C{\isacharprime}{\isacharprime}{\isacharbraceright}{\isacharparenright}\ drl\ in\isanewline
\ \ \ \ \ \ \ensuremath{\underline{\isasymand}}R\ {\isacharparenleft}\ensuremath{{\isasymGamma}_{\isadigit{1}}}{\isacharcomma}\ \ensuremath{{\isasymDelta}_{\isadigit{1}}}\ {\isasymunion}\ {\isacharbraceleft}\ensuremath{\underline{\isasymand}}\ C{\isacharprime}\ C{\isacharprime}{\isacharprime}{\isacharbraceright}{\isacharparenright}\ dll{\isacharprime}\ drl{\isacharprime}\isanewline
{\isachardoublequote}%
\end{isabelle}
}
Similarly

\begin{scriptsize}
\[
\begin{prooftree}
\[
\[
\[
\newover{dlr}{\seq{C',\Gamma_2}{\Delta_2,A}}
\justifies
\seq{C'',C',\Gamma_2}{\Delta_2,A}
\using
\WeakL
\]
\justifies
\seq{C' \land C'',C'',C',\Gamma_2}{\Delta_2,A}
\using
\WeakL
\]
\justifies
\seq{C' \land C'', \Gamma_2}{\Delta_2,A}
\using
{\land L}
\]
\[
\[
\[
\newover{drr}{\seq{C'',\Gamma_2}{\Delta_2,B}}
\justifies
\seq{C',C'',\Gamma_2}{\Delta_2,B}
\using
\WeakL
\]
\justifies
\seq{C' \land C'',C',C'',\Gamma_2}{\Delta_2,B}
\using
\WeakL
\]
\justifies
\seq{C' \land C'',\Gamma_2}{\Delta_2,B}
\using
{\land L}
\]
\justifies
\seq{C' \land C'', \Gamma_2}{\Delta_2}
\using
{\land R}
\end{prooftree}
\]
\end{scriptsize}

The formal proof witness is

\isab{\begin{isabelle}%
{\isachardoublequote}\isanewline
\ \ \ \ let\ dlr{\isacharprime}\ {\isacharequal}\ WL\ {\isacharparenleft}{\isacharbraceleft}C{\isacharprime}{\isacharprime}{\isacharcomma}C{\isacharprime}{\isacharbraceright}\ {\isasymunion}\ \ensuremath{{\isasymGamma}_{\isadigit{2}}}{\isacharcomma}\ensuremath{{\isasymDelta}_{\isadigit{2}}}\ {\isasymunion}\ {\isacharbraceleft}A{\isacharbraceright}{\isacharparenright}\ dlr\ in\isanewline
\ \ \ \ let\ dlr{\isacharprime}{\isacharprime}\ {\isacharequal}\ WL\ {\isacharparenleft}{\isacharbraceleft}\ensuremath{\underline{\isasymand}}\ C{\isacharprime}\ C{\isacharprime}{\isacharprime}{\isacharcomma}C{\isacharprime}{\isacharprime}{\isacharcomma}C{\isacharprime}{\isacharbraceright}\ {\isasymunion}\ \ensuremath{{\isasymGamma}_{\isadigit{2}}}{\isacharcomma}\ \ensuremath{{\isasymDelta}_{\isadigit{2}}}\ {\isasymunion}\ {\isacharbraceleft}A{\isacharbraceright}{\isacharparenright}\ dlr{\isacharprime}\ in\isanewline
\ \ \ \ let\ dlr{\isacharprime}{\isacharprime}{\isacharprime}\ {\isacharequal}\ \ensuremath{\underline{\isasymand}}L\ {\isacharparenleft}{\isacharbraceleft}\ensuremath{\underline{\isasymand}}\ C{\isacharprime}\ C{\isacharprime}{\isacharprime}{\isacharbraceright}\ {\isasymunion}\ \ensuremath{{\isasymGamma}_{\isadigit{2}}}{\isacharcomma}\ \ensuremath{{\isasymDelta}_{\isadigit{2}}}\ {\isasymunion}\ {\isacharbraceleft}A{\isacharbraceright}{\isacharparenright}\ dlr{\isacharprime}{\isacharprime}\ in\ \isanewline
\ \ \ \ \isanewline
\ \ \ \ let\ drr{\isacharprime}\ {\isacharequal}\ WL\ {\isacharparenleft}{\isacharbraceleft}C{\isacharprime}{\isacharcomma}C{\isacharprime}{\isacharprime}{\isacharbraceright}\ {\isasymunion}\ \ensuremath{{\isasymGamma}_{\isadigit{2}}}{\isacharcomma}\ensuremath{{\isasymDelta}_{\isadigit{2}}}\ {\isasymunion}\ {\isacharbraceleft}B{\isacharbraceright}{\isacharparenright}\ drr\ in\isanewline
\ \ \ \ let\ drr{\isacharprime}{\isacharprime}\ {\isacharequal}\ WL\ {\isacharparenleft}{\isacharbraceleft}\ensuremath{\underline{\isasymand}}\ C{\isacharprime}\ C{\isacharprime}{\isacharprime}{\isacharcomma}C{\isacharprime}{\isacharcomma}C{\isacharprime}{\isacharprime}{\isacharbraceright}\ {\isasymunion}\ \ensuremath{{\isasymGamma}_{\isadigit{2}}}{\isacharcomma}\ \ensuremath{{\isasymDelta}_{\isadigit{2}}}\ {\isasymunion}\ {\isacharbraceleft}B{\isacharbraceright}{\isacharparenright}\ drr{\isacharprime}\ in\isanewline
\ \ \ \ let\ drr{\isacharprime}{\isacharprime}{\isacharprime}\ {\isacharequal}\ \ensuremath{\underline{\isasymand}}L\ {\isacharparenleft}{\isacharbraceleft}\ensuremath{\underline{\isasymand}}\ C{\isacharprime}\ C{\isacharprime}{\isacharprime}{\isacharbraceright}\ {\isasymunion}\ \ensuremath{{\isasymGamma}_{\isadigit{2}}}{\isacharcomma}\ \ensuremath{{\isasymDelta}_{\isadigit{2}}}\ {\isasymunion}\ {\isacharbraceleft}B{\isacharbraceright}{\isacharparenright}\ drr{\isacharprime}{\isacharprime}\ in\ \isanewline
\ \ \ \ \isanewline
\ \ \ \ \ \ \ensuremath{\underline{\isasymand}}R\ {\isacharparenleft}{\isacharbraceleft}\ensuremath{\underline{\isasymand}}\ C{\isacharprime}\ C{\isacharprime}{\isacharprime}{\isacharbraceright}\ {\isasymunion}\ \ensuremath{{\isasymGamma}_{\isadigit{2}}}{\isacharcomma}\ensuremath{{\isasymDelta}_{\isadigit{2}}}{\isacharparenright}\ dlr{\isacharprime}{\isacharprime}{\isacharprime}\ drr{\isacharprime}{\isacharprime}{\isacharprime}\isanewline
{\isachardoublequote}%
\end{isabelle}
}

\end{itemize}

\tsubsection{Case $\lor L$} Symmetric to $\land R$.

\tsubsection{Case $\lor R$} Symmetric to $\land L$.

%%%%%%%%%%%%%%%%%%%%%%%%%%%%%%%%%%%%%%%%
\tsubsection{Case $\lnot L$}

We have 

\[
\begin{prooftree}
\newover{d}{\seq{\Gamma_1,\Gamma_2}{\Delta_1,\Delta_2,A}}
\justifies
\seq{\Gamma_1,\Gamma_2}{\Delta_1,\Delta_2}
\using
{\lnot L}
\end{prooftree}
\]

From wellformedness of the derivation, we have that $\lnot A \in
\Gamma_1,\Gamma_2$. There are two subcases, $\lnot A \in \Gamma_1$
or $\lnot A \in \Gamma_2$.

\begin{itemize}

\item Case $\lnot A \in \Gamma_1$. Then the I.H. applied to $d$ gives $C',dl,dr$ such that

\[
\begin{prooftree}
\newover{dl}{\seq{\Gamma_1}{\Delta_1,A,C'}}
\skipand 
\newover{dr}{\seq{C', \Gamma_2}{\Delta_2}}
\thickness0em
\justifies
\end{prooftree}
\]

Take $C = C'$. Then

\[
\begin{prooftree}
\newover{dl}{\seq{\Gamma_1}{\Delta_1,A,C'}}
\justifies
\seq{\Gamma_1}{\Delta_1,C'}
\using
{\lnot L}
\end{prooftree}
\]

The formal proof witness is 

\isab{\begin{isabelle}%
{\isachardoublequote}\isanewline
\ensuremath{\underline{\isasymnot}}L\ {\isacharparenleft}\ensuremath{{\isasymGamma}_{\isadigit{1}}}{\isacharcomma}\ensuremath{{\isasymDelta}_{\isadigit{1}}}\ {\isasymunion}\ {\isacharbraceleft}C{\isacharprime}{\isacharbraceright}{\isacharparenright}\ dl\isanewline
{\isachardoublequote}%
\end{isabelle}
}
$dr$ is already a witness for $\seq{C',\Gamma_2}{\Delta_2}$.

\item Case $\lnot A \in \Gamma_2$. Then the I.H. applied to $d$ gives $C',dl,dr$ such that

\[
\begin{prooftree}
\newover{dl}{\seq{\Gamma_1}{\Delta_1,C'}}
\skipand 
\newover{dr}{\seq{C', \Gamma_2}{\Delta_2,A}}
\thickness0em
\justifies
\end{prooftree}
\]

Take $C = C'$. Then

\[
\begin{prooftree}
\newover{dr}{\seq{C',\Gamma_2}{\Delta_2,A}}
\justifies
\seq{C',\Gamma_2}{\Delta_2}
\using
{\lnot L}
\end{prooftree}
\]

The formal proof witness is 

\isab{\begin{isabelle}%
{\isachardoublequote}\isanewline
\ensuremath{\underline{\isasymnot}}L\ {\isacharparenleft}{\isacharbraceleft}C{\isacharprime}{\isacharbraceright}\ {\isasymunion}\ \ensuremath{{\isasymGamma}_{\isadigit{2}}}{\isacharcomma}\ \ensuremath{{\isasymDelta}_{\isadigit{2}}}{\isacharparenright}\ dr\isanewline
{\isachardoublequote}%
\end{isabelle}
}
$dl$ is already a witness for $\seq{\Gamma_1}{\Delta_1,C'}$.

\end{itemize}

\tsubsection{Case $\lnot R$} Symmetric to $\lnot L$.

%%%%%%%%%%%%%%%%%%%%%%%%%%%%%%%%%%%%%%%%%%%%%%%%%%%%%%%%%%%%%%%%%%%%%%%%%%%%%%%%
\tsubsection{Case $\forall L$}

We have 

\[
\begin{prooftree}
\newover{d}{\seq{A[t],\Gamma_1,\Gamma_2}{\Delta_1,\Delta_2}}
\justifies
\seq{\Gamma_1,\Gamma_2}{\Delta_1,\Delta_2}
\using
{\forall L}
\end{prooftree}
\]

From wellformedness of the derivation, we have that $\forall x. A[x] \in
\Gamma_1,\Gamma_2$. There are two subcases, $\forall x. A[x] \in \Gamma_1$
or $\forall x. A[x] \in \Gamma_2$.

\begin{itemize}

\item Case $\forall x. A[x] \in \Gamma_1$. Then the I.H. applied to $d$ gives $C',dl,dr$ such that

\[
\begin{prooftree}
\newover{dl}{\seq{A[t],\Gamma_1}{\Delta_1,C'}}
\skipand 
\newover{dr}{\seq{C', \Gamma_2}{\Delta_2}}
\thickness0em
\justifies
\end{prooftree}
\]

Take $C = C'$. Then

\[
\begin{prooftree}
\newover{dl}{\seq{A[t],\Gamma_1}{\Delta_1,C'}}
\justifies
\seq{\Gamma_1}{\Delta_1,C'}
\using
{\forall L}
\end{prooftree}
\]

The formal proof witness is 

\isab{\begin{isabelle}%
{\isachardoublequote}\isanewline
\ensuremath{\underline{{\isasymforall}}}L\ {\isacharparenleft}\ensuremath{{\isasymGamma}_{\isadigit{1}}}{\isacharcomma}\ \ensuremath{{\isasymDelta}_{\isadigit{1}}}\ {\isasymunion}\ {\isacharbraceleft}C{\isacharprime}{\isacharbraceright}{\isacharparenright}\ dl\isanewline
{\isachardoublequote}%
\end{isabelle}
}
$dr$ is already a witness for $\seq{C',\Gamma_2}{\Delta_2}$.

\item Case $\forall x. A[x] \in \Gamma_2$. Then the I.H. applied to $d$ gives $C',dl,dr$ such that

\[
\begin{prooftree}
\newover{dl}{\seq{\Gamma_1}{\Delta_1,C'}}
\skipand 
\newover{dr}{\seq{C', A[t],\Gamma_2}{\Delta_2}}
\thickness0em
\justifies
\end{prooftree}
\]

Take $C = C'$. Then

\[
\begin{prooftree}
\newover{dr}{\seq{C',A[t],\Gamma_2}{\Delta_2}}
\justifies
\seq{C',\Gamma_2}{\Delta_2}
\using
{\forall L}
\end{prooftree}
\]

The formal proof witness is 

\isab{\begin{isabelle}%
{\isachardoublequote}\isanewline
\ensuremath{\underline{{\isasymforall}}}L\ {\isacharparenleft}{\isacharbraceleft}C{\isacharprime}{\isacharbraceright}\ {\isasymunion}\ \ensuremath{{\isasymGamma}_{\isadigit{2}}}{\isacharcomma}\ \ensuremath{{\isasymDelta}_{\isadigit{2}}}{\isacharparenright}\ dr\isanewline
{\isachardoublequote}%
\end{isabelle}
}
$dl$ is already a witness for $\seq{\Gamma_1}{\Delta_1,C'}$.

\end{itemize}

%%%%%%%%%%%%%%%%%%%%%%%%%%%%%%%%%%%%%%%%%%%%%%%%%%%%%%%%%%%%%%%%%%%%%%%%%%%%%%%%
\tsubsection{Case $\forall R$}

We have 

\[
\begin{prooftree}
\newover{d}{\seq{\Gamma_1,\Gamma_2}{\Delta_1,\Delta_2,A[a]}}
\justifies
\seq{\Gamma_1,\Gamma_2}{\Delta_1,\Delta_2}
\using
{\forall R}
\end{prooftree}
\]

From wellformedness of the derivation, we have that $\forall x. A[x] \in
\Delta_1,\Delta_2$. There are two subcases, $\forall x. A[x] \in \Delta_1$
or $\forall x. A[x] \in \Delta_2$.

\begin{itemize}

%%%%%%%%%%%%%%%%%%%%%%%%%%%%%%%%%%%%%%%%%%%%%%%%%%%%%%%%%%%%%%%%%%%%%%%%%%%%%%%%
\item Case $\forall x. A[x] \in \Delta_1$. Then the I.H. applied to $d$ gives $C'[a],dl,dr$ such that

\[
\begin{prooftree}
\newover{dl}{\seq{\Gamma_1}{\Delta_1,A[a],C'[a]}}
\skipand 
\newover{dr}{\seq{C'[a], \Gamma_2}{\Delta_2}}
\thickness0em
\justifies
\end{prooftree}
\]

Take $C = \exists x. C'[x]$. Then

\[
\begin{prooftree}
\[
\[
\newover{dl}{\seq{\Gamma_1}{\Delta_1,A[a],C'[a]}}
\justifies
\seq{\Gamma_1}{\Delta_1,A[a],C'[a],\exists x. C'[x]}
\using
\WeakR
\]
\justifies
\seq{\Gamma_1}{\Delta_1,A[a],\exists x. C'[x]}
\using
{\exists R}
\]
\justifies
\seq{\Gamma_1}{\Delta_1,\exists x. C'[x]}
\using
{\forall R}
\end{prooftree}
\]

The formal proof witness is 

\isab{\begin{isabelle}%
{\isachardoublequote}\isanewline
\ \ \ \ let\ dl{\isacharprime}\ {\isacharequal}\ WR\ {\isacharparenleft}\ensuremath{{\isasymGamma}_{\isadigit{1}}}{\isacharcomma}\ \ensuremath{{\isasymDelta}_{\isadigit{1}}}\ {\isasymunion}\ {\isacharbraceleft}A{\isacharcomma}\ C{\isacharprime}{\isacharcomma}\ \ensuremath{\underline{\isasymexists}}\ a\ C{\isacharprime}{\isacharbraceright}{\isacharparenright}\ dl\ in\isanewline
\ \ \ \ let\ dl{\isacharprime}{\isacharprime}\ {\isacharequal}\ \ensuremath{\underline{{\isasymexists}}}R\ {\isacharparenleft}\ensuremath{{\isasymGamma}_{\isadigit{1}}}{\isacharcomma}\ensuremath{{\isasymDelta}_{\isadigit{1}}}\ {\isasymunion}\ {\isacharbraceleft}A{\isacharcomma}\ \ensuremath{\underline{\isasymexists}}\ a\ C{\isacharprime}{\isacharbraceright}{\isacharparenright}\ dl{\isacharprime}\ in\isanewline
\ \ \ \ \ \ \ensuremath{\underline{{\isasymforall}}}R\ {\isacharparenleft}\ensuremath{{\isasymGamma}_{\isadigit{1}}}{\isacharcomma}\ensuremath{{\isasymDelta}_{\isadigit{1}}}\ {\isasymunion}\ {\isacharbraceleft}\ensuremath{\underline{\isasymexists}}\ a\ C{\isacharprime}{\isacharbraceright}{\isacharparenright}\ dl{\isacharprime}{\isacharprime}\isanewline
{\isachardoublequote}%
\end{isabelle}
}
Similarly

\[
\begin{prooftree}
\[
\newover{dr}{\seq{C'[a],\Gamma_2}{\Delta_2}}
\justifies
\seq{\exists x. C'[x], C'[a], \Gamma_2}{\Delta_2}
\using
\WeakL
\]
\justifies
\seq{\exists x. C'[x], \Gamma_2}{\Delta_2}
\using
{\exists L}
\end{prooftree}
\]

The formal proof witness is 

\isab{\begin{isabelle}%
{\isachardoublequote}\isanewline
\ \ \ \ let\ dr{\isacharprime}\ {\isacharequal}\ WL\ {\isacharparenleft}{\isacharbraceleft}\ensuremath{\underline{\isasymexists}}\ a\ C{\isacharprime}{\isacharbraceright}\ {\isasymunion}\ {\isacharbraceleft}C{\isacharprime}{\isacharbraceright}\ {\isasymunion}\ \ensuremath{{\isasymGamma}_{\isadigit{2}}}{\isacharcomma}\ \ensuremath{{\isasymDelta}_{\isadigit{2}}}{\isacharparenright}\ dr\ in\isanewline
\ \ \ \ \ \ \ensuremath{\underline{{\isasymexists}}}L\ {\isacharparenleft}{\isacharbraceleft}\ensuremath{\underline{\isasymexists}}\ a\ C{\isacharprime}{\isacharbraceright}\ {\isasymunion}\ \ensuremath{{\isasymGamma}_{\isadigit{2}}}{\isacharcomma}\ensuremath{{\isasymDelta}_{\isadigit{2}}}{\isacharparenright}\ dr{\isacharprime}\isanewline
{\isachardoublequote}%
\end{isabelle}
}
%%%%%%%%%%%%%%%%%%%%%%%%%%%%%%%%%%%%%%%%%%%%%%%%%%%%%%%%%%%%%%%%%%%%%%%%%%%%%%%%
\item Case $\forall x. A[x] \in \Delta_2$. Then the I.H. applied to $d$ gives $C'[a],dl,dr$ such that

\[
\begin{prooftree}
\newover{dl}{\seq{\Gamma_1}{\Delta_1,C'[a]}}
\skipand 
\newover{dr}{\seq{C'[a], \Gamma_2}{\Delta_2,A[a]}}
\thickness0em
\justifies
\end{prooftree}
\]

Take $C = \forall x. C'[x]$. Then

\[
\begin{prooftree}
\[
\newover{dl}{\seq{\Gamma_1}{\Delta_1,C'[a]}}
\justifies
\seq{\Gamma_1}{\Delta_1,C'[a],\forall x. C'[x]}
\using
\WeakR
\]
\justifies
\seq{\Gamma_1}{\Delta_1,\forall x. C'[x]}
\using
{\forall R}
\end{prooftree}
\]

The formal proof witness is 

\isab{\begin{isabelle}%
{\isachardoublequote}\isanewline
\ \ \ \ let\ dl{\isacharprime}\ {\isacharequal}\ WR\ {\isacharparenleft}\ensuremath{{\isasymGamma}_{\isadigit{1}}}{\isacharcomma}\ \ensuremath{{\isasymDelta}_{\isadigit{1}}}\ {\isasymunion}\ {\isacharbraceleft}C{\isacharprime}{\isacharcomma}\ensuremath{\underline{\isasymforall}}\ a\ C{\isacharprime}{\isacharbraceright}{\isacharparenright}\ dl\ in\isanewline
\ \ \ \ \ \ \ensuremath{\underline{{\isasymforall}}}R\ {\isacharparenleft}\ensuremath{{\isasymGamma}_{\isadigit{1}}}{\isacharcomma}\ \ensuremath{{\isasymDelta}_{\isadigit{1}}}\ {\isasymunion}\ {\isacharbraceleft}\ensuremath{\underline{\isasymforall}}\ a\ C{\isacharprime}{\isacharbraceright}{\isacharparenright}\ dl{\isacharprime}\isanewline
{\isachardoublequote}%
\end{isabelle}
}
Similarly

\[
\begin{prooftree}
\[
\[
\newover{dr}{\seq{C'[a],\Gamma_2}{\Delta_2,A[a]}}
\justifies
\seq{\forall x. C'[x], C'[a], \Gamma_2}{\Delta_2,A[a]}
\using
\WeakL
\]
\justifies
\seq{\forall x. C'[x], \Gamma_2}{\Delta_2,A[a]}
\using
{\forall L}
\]
\justifies
\seq{\forall x. C'[x], \Gamma_2}{\Delta_2}
\using
{\forall R}
\end{prooftree}
\]

The formal proof witness is 

\isab{\begin{isabelle}%
{\isachardoublequote}\isanewline
\ \ \ \ let\ dr{\isacharprime}\ {\isacharequal}\ WL\ {\isacharparenleft}{\isacharbraceleft}\ensuremath{\underline{\isasymforall}}\ a\ C{\isacharprime}{\isacharcomma}C{\isacharprime}{\isacharbraceright}\ {\isasymunion}\ \ensuremath{{\isasymGamma}_{\isadigit{2}}}{\isacharcomma}\ \ensuremath{{\isasymDelta}_{\isadigit{2}}}\ {\isasymunion}\ {\isacharbraceleft}A{\isacharbraceright}{\isacharparenright}\ dr\ in\isanewline
\ \ \ \ let\ dr{\isacharprime}{\isacharprime}\ {\isacharequal}\ \ensuremath{\underline{{\isasymforall}}}L\ {\isacharparenleft}{\isacharbraceleft}\ensuremath{\underline{\isasymforall}}\ a\ C{\isacharprime}{\isacharbraceright}\ {\isasymunion}\ \ensuremath{{\isasymGamma}_{\isadigit{2}}}{\isacharcomma}\ \ensuremath{{\isasymDelta}_{\isadigit{2}}}\ {\isasymunion}\ {\isacharbraceleft}A{\isacharbraceright}{\isacharparenright}\ dr{\isacharprime}\ in\ \isanewline
\ \ \ \ \ \ \ensuremath{\underline{{\isasymforall}}}R\ {\isacharparenleft}{\isacharbraceleft}\ensuremath{\underline{\isasymforall}}\ a\ C{\isacharprime}{\isacharbraceright}\ {\isasymunion}\ \ensuremath{{\isasymGamma}_{\isadigit{2}}}{\isacharcomma}\ \ensuremath{{\isasymDelta}_{\isadigit{2}}}{\isacharparenright}\ dr{\isacharprime}{\isacharprime}\isanewline
{\isachardoublequote}%
\end{isabelle}
}
\end{itemize}
%%%%%%%%%%%%%%%%%%%%%%%%%%%%%%%%%%%%%%%%%%%%%%%%%%%%%%%%%%%%%%%%%%%%%%%%%%%%%%%%

\tsubsection{Case $\exists L$} Symmetric to $\forall R$.

\tsubsection{Case $\exists R$} Symmetric to $\forall L$.

\tsubsection{Case $\WeakL$}

We have 

\[
\begin{prooftree}
\newover{d}{\seq{\Gamma}{\Delta_1,\Delta_2}}
\justifies
\seq{\Gamma_1,\Gamma_2}{\Delta_1,\Delta_2}
\using
\WeakL
\end{prooftree}
\]

From wellformedness of the derivation we have that $A,\Gamma = \Gamma_1,\Gamma_2$. The I.H. applied to $d$ gives $C',dl,dr$ such that

\[
\begin{prooftree}
\newover{dl}{\seq{\Gamma \cap \Gamma_1}{\Delta_1,C'}}
\skipand 
\newover{dr}{\seq{C', \Gamma \cap \Gamma_2}{\Delta_2}}
\thickness0em
\justifies
\end{prooftree}
\]

Take $C = C'$. There are four subcases.

\begin{itemize}

\item Case $A \in \Gamma_1, A \in \Gamma_2$. Then

\[
\begin{prooftree}
\newover{dl}{\seq{\Gamma \cap \Gamma_1}{\Delta_1,C'}}
\justifies
\seq{\Gamma_1}{\Delta_1,C'}
\using
\WeakL
\end{prooftree}
\]

The formal proof witness is 

\isab{\begin{isabelle}%
{\isachardoublequote}\isanewline
WL\ {\isacharparenleft}\ensuremath{{\isasymGamma}_{\isadigit{1}}}{\isacharcomma}\ \ensuremath{{\isasymDelta}_{\isadigit{1}}}\ {\isasymunion}\ {\isacharbraceleft}C{\isacharprime}{\isacharbraceright}{\isacharparenright}\ dl\isanewline
{\isachardoublequote}%
\end{isabelle}
}
Similarly

\[
\begin{prooftree}
\newover{dr}{\seq{C',\Gamma \cap \Gamma_2}{\Delta_2}}
\justifies
\seq{C',\Gamma_2}{\Delta_2}
\using
\WeakL
\end{prooftree}
\]

The formal proof witness is 

\isab{\begin{isabelle}%
{\isachardoublequote}\isanewline
WL\ {\isacharparenleft}{\isacharbraceleft}C{\isacharprime}{\isacharbraceright}\ {\isasymunion}\ \ensuremath{{\isasymGamma}_{\isadigit{2}}}{\isacharcomma}\ \ensuremath{{\isasymDelta}_{\isadigit{2}}}{\isacharparenright}\ dr\isanewline
{\isachardoublequote}%
\end{isabelle}
}
\item Case $A \in \Gamma_1, A \notin \Gamma_2$. Then

\[
\begin{prooftree}
\newover{dl}{\seq{\Gamma \cap \Gamma_1}{\Delta_1,C'}}
\justifies
\seq{\Gamma_1}{\Delta_1,C'}
\using
\WeakL
\end{prooftree}
\]

The formal proof witness is 

\isab{\begin{isabelle}%
{\isachardoublequote}\isanewline
WL\ {\isacharparenleft}\ensuremath{{\isasymGamma}_{\isadigit{1}}}{\isacharcomma}\ \ensuremath{{\isasymDelta}_{\isadigit{1}}}\ {\isasymunion}\ {\isacharbraceleft}C{\isacharprime}{\isacharbraceright}{\isacharparenright}\ dl\isanewline
{\isachardoublequote}%
\end{isabelle}
}
$dr$ is already a witness for $\seq{C',\Gamma_2}{\Delta_2}$.

\item Case $A \notin \Gamma_1, A \in \Gamma_2$. Symmetric to previous case.

\item Case $A \notin \Gamma_1, A \notin \Gamma_2$. Contradiction.

\end{itemize}

%%%%%%%%%%%%%%%%%%%%%%%%%%%%%%%%%%%%%%%%%%%%%%%%%%%%%%%%%%%%%%%%%%%%%%%%%%%%%%%%
\tsubsection{Case $\WeakR$} Symmetric to $\WeakL$.

%%%%%%%%%%%%%%%%%%%%%%%%%%%%%%%%%%%%%%%%%%%%%%%%%%%%%%%%%%%%%%%%%%%%%%%%%%%%%%%%
\tsection{Concrete Development of Formulae}

\label{sectCraigConcrete}

The development described in the previous sections is axiomatic. We
also provide a fully conservative definition based on a de Bruijn
representation of binding 
%cite{FIXME} 
for formulae.

Formulae are defined as follows.

\isab{\isanewline
\isacommand{datatype}\isamarkupfalse%
\ form\ {\isacharequal}\ P\ pred\ {\isachardoublequoteopen}{\isacharparenleft}tm\ list{\isacharparenright}{\isachardoublequoteclose}\ \ \isanewline
\ \ {\isasymor}\ \ensuremath{\underline{\isasymbottom}}\isanewline
\ \ {\isasymor}\ \ensuremath{\underline{\isasymtop}}\isanewline
\ \ {\isasymor}\ \ensuremath{\underline{\isasymand}}\ form\ form\ \isanewline
\ \ {\isasymor}\ \ensuremath{\underline{\isasymor}}\ form\ form\isanewline
\ \ {\isasymor}\ \ensuremath{\underline{\isasymnot}}\ form\ \isanewline
\ \ {\isasymor}\ FAll\ form\isanewline
\ \ {\isasymor}\ FEx\ form\isanewline
}
Substitution is defined as usual.

\isab{\isanewline
\isacommand{consts}\isamarkupfalse%
\ fsubst\ {\isacharcolon}{\isacharcolon}\ {\isachardoublequoteopen}{\isacharparenleft}var\ {\isasymRightarrow}\ tm{\isacharparenright}\ {\isasymRightarrow}\ form\ {\isasymRightarrow}\ form{\isachardoublequoteclose}\isanewline
\isacommand{primrec}\isamarkupfalse%
\ \isanewline
\ \ {\isachardoublequoteopen}fsubst\ s\ {\isacharparenleft}P\ i\ tms{\isacharparenright}\ {\isacharequal}\ P\ i\ {\isacharparenleft}map\ s\ tms{\isacharparenright}{\isachardoublequoteclose}\isanewline
\ \ {\isachardoublequoteopen}fsubst\ s\ \ensuremath{\underline{\isasymbottom}}\ {\isacharequal}\ \ensuremath{\underline{\isasymbottom}}{\isachardoublequoteclose}\isanewline
\ \ {\isachardoublequoteopen}fsubst\ s\ \ensuremath{\underline{\isasymtop}}\ {\isacharequal}\ \ensuremath{\underline{\isasymtop}}{\isachardoublequoteclose}\isanewline
\ \ {\isachardoublequoteopen}fsubst\ s\ {\isacharparenleft}\ensuremath{\underline{\isasymand}}\ A\ B{\isacharparenright}\ {\isacharequal}\ \ensuremath{\underline{\isasymand}}\ {\isacharparenleft}fsubst\ s\ A{\isacharparenright}\ {\isacharparenleft}fsubst\ s\ B{\isacharparenright}{\isachardoublequoteclose}\isanewline
\ \ {\isachardoublequoteopen}fsubst\ s\ {\isacharparenleft}\ensuremath{\underline{\isasymor}}\ A\ B{\isacharparenright}\ {\isacharequal}\ \ensuremath{\underline{\isasymor}}\ {\isacharparenleft}fsubst\ s\ A{\isacharparenright}\ {\isacharparenleft}fsubst\ s\ B{\isacharparenright}{\isachardoublequoteclose}\isanewline
\ \ {\isachardoublequoteopen}fsubst\ s\ {\isacharparenleft}\ensuremath{\underline{\isasymnot}}\ A{\isacharparenright}\ {\isacharequal}\ \ensuremath{\underline{\isasymnot}}\ {\isacharparenleft}fsubst\ s\ A{\isacharparenright}{\isachardoublequoteclose}\isanewline
\ \ {\isachardoublequoteopen}fsubst\ s\ {\isacharparenleft}FAll\ A{\isacharparenright}\ {\isacharequal}\ {\isacharparenleft}let\ s\ {\isacharequal}\ {\isacharparenleft}{\isasymlambda}\ v{\isachardot}\ case\ v\ of\ {\isadigit{0}}\ {\isasymRightarrow}\ {\isadigit{0}}\ {\isasymor}\ Suc\ n\ {\isasymRightarrow}\ Suc\ {\isacharparenleft}s\ n{\isacharparenright}{\isacharparenright}\ in\ FAll\ {\isacharparenleft}fsubst\ s\ A{\isacharparenright}{\isacharparenright}{\isachardoublequoteclose}\isanewline
\ \ {\isachardoublequoteopen}fsubst\ s\ {\isacharparenleft}FEx\ A{\isacharparenright}\ {\isacharequal}\ {\isacharparenleft}let\ s\ {\isacharequal}\ {\isacharparenleft}{\isasymlambda}\ v{\isachardot}\ case\ v\ of\ {\isadigit{0}}\ {\isasymRightarrow}\ {\isadigit{0}}\ {\isasymor}\ Suc\ n\ {\isasymRightarrow}\ Suc\ {\isacharparenleft}s\ n{\isacharparenright}{\isacharparenright}\ in\ FEx\ {\isacharparenleft}fsubst\ s\ A{\isacharparenright}{\isacharparenright}{\isachardoublequoteclose}\isanewline
}
The axiomatic formulae constructors are defined concretely as follows.

\isab{\isanewline
\isacommand{consts}\isamarkupfalse%
\isanewline
\ \ \ensuremath{\underline{\isasymforall}}\ {\isacharcolon}{\isacharcolon}\ {\isachardoublequoteopen}var\ {\isasymRightarrow}\ form\ {\isasymRightarrow}\ form{\isachardoublequoteclose}\isanewline
\ \ \ensuremath{\underline{\isasymexists}}\ {\isacharcolon}{\isacharcolon}\ {\isachardoublequoteopen}var\ {\isasymRightarrow}\ form\ {\isasymRightarrow}\ form{\isachardoublequoteclose}\isanewline
\isanewline
\isacommand{defs}\isamarkupfalse%
\ \isanewline
\ \ {\isachardoublequoteopen}\ensuremath{\underline{\isasymforall}}\ a\ A\ {\isasymequiv}\ FAll\ {\isacharparenleft}fsubst\ {\isacharparenleft}{\isasymlambda}\ v{\isachardot}\ if\ v\ {\isacharequal}\ a\ then\ {\isadigit{0}}\ else\ Suc\ v{\isacharparenright}\ A{\isacharparenright}{\isachardoublequoteclose}\isanewline
\ \ {\isachardoublequoteopen}\ensuremath{\underline{\isasymexists}}\ a\ A\ {\isasymequiv}\ FEx\ {\isacharparenleft}fsubst\ {\isacharparenleft}{\isasymlambda}\ v{\isachardot}\ if\ v\ {\isacharequal}\ a\ then\ {\isadigit{0}}\ else\ Suc\ v{\isacharparenright}\ A{\isacharparenright}{\isachardoublequoteclose}\isanewline
}
Instantiation of quantified formulae is defined as follows.

\isab{%
\endisatagproof
{\isafoldproof}%
\isadelimproof
\isanewline
\endisadelimproof
\isacommand{consts}\isamarkupfalse%
\isanewline
\ \ FAll{\isacharunderscore}inst\ {\isacharcolon}{\isacharcolon}\ {\isachardoublequoteopen}tm\ {\isasymRightarrow}\ form\ {\isasymRightarrow}\ form{\isachardoublequoteclose}\isanewline
\ \ FEx{\isacharunderscore}inst\ {\isacharcolon}{\isacharcolon}\ {\isachardoublequoteopen}tm\ {\isasymRightarrow}\ form\ {\isasymRightarrow}\ form{\isachardoublequoteclose}\isanewline
\isacommand{primrec}\isamarkupfalse%
\isanewline
\ \ {\isachardoublequoteopen}FAll{\isacharunderscore}inst\ t\ {\isacharparenleft}FAll\ A{\isacharparenright}\ {\isacharequal}\ fsubst\ {\isacharparenleft}{\isasymlambda}\ v{\isachardot}\ case\ v\ of\ {\isadigit{0}}\ {\isasymRightarrow}\ t\ {\isasymor}\ Suc\ n\ {\isasymRightarrow}\ n{\isacharparenright}\ A{\isachardoublequoteclose}\isanewline
\isacommand{primrec}\isamarkupfalse%
\isanewline
\ \ {\isachardoublequoteopen}FEx{\isacharunderscore}inst\ t\ {\isacharparenleft}FEx\ A{\isacharparenright}\ {\isacharequal}\ fsubst\ {\isacharparenleft}{\isasymlambda}\ v{\isachardot}\ case\ v\ of\ {\isadigit{0}}\ {\isasymRightarrow}\ t\ {\isasymor}\ Suc\ n\ {\isasymRightarrow}\ n{\isacharparenright}\ A{\isachardoublequoteclose}\isanewline
}
Positive and negative occurrences are defined in a mutually recursive fashion.

\isab{
\isanewline
\isacommand{consts}\isamarkupfalse%
\ posneg\ {\isacharcolon}{\isacharcolon}\ {\isachardoublequoteopen}form\ {\isasymRightarrow}\ pred\ set\ {\isacharasterisk}\ pred\ set{\isachardoublequoteclose}\isanewline
\isacommand{primrec}\isamarkupfalse%
\ \isanewline
\ \ {\isachardoublequoteopen}posneg\ {\isacharparenleft}P\ i\ vs{\isacharparenright}\ {\isacharequal}\ {\isacharparenleft}{\isacharbraceleft}i{\isacharbraceright}{\isacharcomma}{\isacharbraceleft}{\isacharbraceright}{\isacharparenright}{\isachardoublequoteclose}\isanewline
\ \ {\isachardoublequoteopen}posneg\ \ensuremath{\underline{\isasymbottom}}\ {\isacharequal}\ {\isacharparenleft}{\isacharbraceleft}{\isacharbraceright}{\isacharcomma}{\isacharbraceleft}{\isacharbraceright}{\isacharparenright}{\isachardoublequoteclose}\isanewline
\ \ {\isachardoublequoteopen}posneg\ \ensuremath{\underline{\isasymtop}}\ {\isacharequal}\ {\isacharparenleft}{\isacharbraceleft}{\isacharbraceright}{\isacharcomma}{\isacharbraceleft}{\isacharbraceright}{\isacharparenright}{\isachardoublequoteclose}\isanewline
\ \ {\isachardoublequoteopen}posneg\ {\isacharparenleft}\ensuremath{\underline{\isasymand}}\ f\ g{\isacharparenright}\ {\isacharequal}\ {\isacharparenleft}let\ {\isacharparenleft}fp{\isacharcomma}fn{\isacharparenright}\ {\isacharequal}\ posneg\ f\ in\ \isanewline
\ \ \ \ let\ {\isacharparenleft}gp{\isacharcomma}gn{\isacharparenright}\ {\isacharequal}\ posneg\ g\ in\isanewline
\ \ \ \ {\isacharparenleft}fp\ {\isasymunion}\ gp{\isacharcomma}\ fn\ {\isasymunion}\ gn{\isacharparenright}{\isacharparenright}{\isachardoublequoteclose}\isanewline
\ \ {\isachardoublequoteopen}posneg\ {\isacharparenleft}\ensuremath{\underline{\isasymor}}\ f\ g{\isacharparenright}\ {\isacharequal}\ {\isacharparenleft}let\ {\isacharparenleft}fp{\isacharcomma}fn{\isacharparenright}\ {\isacharequal}\ posneg\ f\ in\ \isanewline
\ \ \ \ let\ {\isacharparenleft}gp{\isacharcomma}gn{\isacharparenright}\ {\isacharequal}\ posneg\ g\ in\isanewline
\ \ \ \ {\isacharparenleft}fp\ {\isasymunion}\ gp{\isacharcomma}\ fn\ {\isasymunion}\ gn{\isacharparenright}{\isacharparenright}{\isachardoublequoteclose}\isanewline
\ \ {\isachardoublequoteopen}posneg\ {\isacharparenleft}\ensuremath{\underline{\isasymnot}}\ f{\isacharparenright}\ {\isacharequal}\ {\isacharparenleft}let\ {\isacharparenleft}p{\isacharcomma}n{\isacharparenright}\ {\isacharequal}\ posneg\ f\ in\ {\isacharparenleft}n{\isacharcomma}p{\isacharparenright}{\isacharparenright}{\isachardoublequoteclose}\isanewline
\ \ {\isachardoublequoteopen}posneg\ {\isacharparenleft}FAll\ f{\isacharparenright}\ {\isacharequal}\ posneg\ f{\isachardoublequoteclose}\isanewline
\ \ {\isachardoublequoteopen}posneg\ {\isacharparenleft}FEx\ f{\isacharparenright}\ {\isacharequal}\ posneg\ f{\isachardoublequoteclose}\isanewline
\isanewline
\isacommand{constdefs}\isamarkupfalse%
\ pos\ {\isacharcolon}{\isacharcolon}\ {\isachardoublequoteopen}form\ {\isasymRightarrow}\ pred\ set{\isachardoublequoteclose}\isanewline
\ \ {\isachardoublequoteopen}pos\ {\isasymequiv}\ fst\ o\ posneg{\isachardoublequoteclose}\isanewline
\isanewline
\isacommand{constdefs}\isamarkupfalse%
\ neg\ {\isacharcolon}{\isacharcolon}\ {\isachardoublequoteopen}form\ {\isasymRightarrow}\ pred\ set{\isachardoublequoteclose}\isanewline
\ \ {\isachardoublequoteopen}neg\ {\isasymequiv}\ snd\ o\ posneg{\isachardoublequoteclose}\isanewline
}

Free variables are defined using an auxiliary function.

\isab{
\endisatagproof
{\isafoldproof}%
\isadelimproof
\isanewline
\endisadelimproof
\isacommand{consts}\isamarkupfalse%
\ preSuc\ {\isacharcolon}{\isacharcolon}\ {\isachardoublequoteopen}nat\ list\ {\isasymRightarrow}\ nat\ list{\isachardoublequoteclose}\isanewline
\isacommand{primrec}\isamarkupfalse%
\isanewline
\ \ {\isachardoublequoteopen}preSuc\ {\isacharbrackleft}{\isacharbrackright}\ {\isacharequal}\ {\isacharbrackleft}{\isacharbrackright}{\isachardoublequoteclose}\isanewline
\ \ {\isachardoublequoteopen}preSuc\ {\isacharparenleft}a{\isacharhash}list{\isacharparenright}\ {\isacharequal}\ {\isacharparenleft}case\ a\ of\ {\isadigit{0}}\ {\isasymRightarrow}\ preSuc\ list\ {\isasymor}\ Suc\ n\ {\isasymRightarrow}\ n{\isacharhash}{\isacharparenleft}preSuc\ list{\isacharparenright}{\isacharparenright}{\isachardoublequoteclose}\isanewline
\ \ %
\isanewline
\isacommand{consts}\isamarkupfalse%
\ fv\ {\isacharcolon}{\isacharcolon}\ {\isachardoublequoteopen}form\ {\isasymRightarrow}\ var\ list{\isachardoublequoteclose}\isanewline
\isacommand{primrec}\isamarkupfalse%
\ \isanewline
\ \ {\isachardoublequoteopen}fv\ {\isacharparenleft}P\ i\ tms{\isacharparenright}\ {\isacharequal}\ tms{\isachardoublequoteclose}\isanewline
\ \ {\isachardoublequoteopen}fv\ \ensuremath{\underline{\isasymbottom}}\ {\isacharequal}\ {\isacharbrackleft}{\isacharbrackright}{\isachardoublequoteclose}\isanewline
\ \ {\isachardoublequoteopen}fv\ \ensuremath{\underline{\isasymtop}}\ {\isacharequal}\ {\isacharbrackleft}{\isacharbrackright}{\isachardoublequoteclose}\isanewline
\ \ {\isachardoublequoteopen}fv\ {\isacharparenleft}\ensuremath{\underline{\isasymand}}\ A\ B{\isacharparenright}\ {\isacharequal}\ {\isacharparenleft}fv\ A{\isacharparenright}\ {\isacharat}\ {\isacharparenleft}fv\ B{\isacharparenright}{\isachardoublequoteclose}\isanewline
\ \ {\isachardoublequoteopen}fv\ {\isacharparenleft}\ensuremath{\underline{\isasymor}}\ A\ B{\isacharparenright}\ {\isacharequal}\ {\isacharparenleft}fv\ A{\isacharparenright}\ {\isacharat}\ {\isacharparenleft}fv\ B{\isacharparenright}{\isachardoublequoteclose}\isanewline
\ \ {\isachardoublequoteopen}fv\ {\isacharparenleft}\ensuremath{\underline{\isasymnot}}\ A{\isacharparenright}\ {\isacharequal}\ fv\ A{\isachardoublequoteclose}\isanewline
\ \ {\isachardoublequoteopen}fv\ {\isacharparenleft}FAll\ A{\isacharparenright}\ {\isacharequal}\ preSuc\ {\isacharparenleft}fv\ A{\isacharparenright}{\isachardoublequoteclose}\isanewline
\ \ {\isachardoublequoteopen}fv\ {\isacharparenleft}FEx\ A{\isacharparenright}\ {\isacharequal}\ preSuc\ {\isacharparenleft}fv\ A{\isacharparenright}{\isachardoublequoteclose}\isanewline
\ \ %
}

All properties which we previously asserted axiomatically are proved
for the corresponding concrete development. The main proof of Craig's
Interpolation Theorem can run happily using either the axiomatic
development or the concrete development.

%%%%%%%%%%%%%%%%%%%%%%%%%%%%%%%%%%%%%%%%%%%%%%%%%%%%%%%%%%%%%%%%%%%%%%%%%%%%%%%%
\tsection{Analysis}

\label{sectCraigAnalysis}

\tsubsection{Formal v. Informal}

In the preceeding sections we have given an informal account of a
formal mechanised proof. We have omitted numerous checks from the
informal proof. For example:

\begin{itemize}
\item We noted already the omission of the checks on the
polarity of predicates appearing in the interpolation formula.
\item  We omitted checking wellformedness of intermediate derivations which
are used as witnesses in the proof. 
\item We omitted cases where symmetry is sufficient to allow the reader to reconstruct the proof
from a previous case. 
\item We omitted eigenvariable checks in
$\forall R, \exists L$ cases. 
\end{itemize}

Suffice it to say, including these details would have substantially
increased the size of the informal presentation.
Never-the-less, the informal presentation is by no means short.

The formal, mechanised version can be significantly shorter than an
informal presentation because much of the proof can be relegated to
automation. However, the formal proof is certainly less readable. 

Ideally one would like the formal and the informal presentation to
inhabit the same document. Ideally the formal terms should be typeset
as informal practice. For example, derivations used in the proof
should be typeset as such, not just quoted as HOL terms. Although
Isabelle possesses some facilities in this area, improvements can
certainly be made.

\tsubsection{Mechanisation Statistics}

Our abstract development of formulae consists of 95 lines (including
whitespace), of which none are tactic lines, and our concrete
development contains 210 lines, of which 51 are tactic lines.

Our main mechanised theory file contains 410 tactic lines.  Each case
in the main proof requires us to prove about 10 different subgoals,
and each subgoal corresponds roughly to a single line of tactic
script. We have 5 connectives or quantifiers, 10 corresponding left
and right rules, and 2 subcases per rule, giving 20 cases in total. In
addition, there are 4 cases for the {\Init} rule, and 4 cases each for
the two {\Weak} rules, giving a total of $20 + 4 + (4 + 4) = 32$
cases in all. At approximately 10 lines per case, this gives rise to
approximately 320 tactic lines, with the rest related to setting up
outside inductions, and the derivation of the weak form of Craig's
Interpolation Theorem.

The total line count is under 1000 lines, and this includes many
whitespace lines, lemmas that reproduce in Isar what previously was
conducted using tactics, and lines whose sole purpose is to requote
formal witnesses so that they can be included in this informal
presentation.

The point is simply that this development is extremely short.

\tsubsection{Aims of the Mechanisation}

In this section, we discuss what we tried to achieve with the
mechanisation. Some of these achievements are far from obvious even
when replaying the mechanised text step by step.

\begin{itemize}
  
\item \textbf{Clear, Correct and Complete} 
  We hope our presentation is clear. Existing presentations are
  lacking in this area. For example, Girard in \cite{girard87proof}
  rephrases the induction statement halfway through the proof, whereas
  we have been careful to state our theorems precisely.
  Moreover, because we have formalised the proofs, many details that
  were murky have been uncovered. A particular area of concern is the
  informal tradition of requiring that the analysed formula appear
  explicitly in the conclusion of a rule. We believe the resulting
  proofs are often hard to read. For example, Girard follows the
  tradition, but the individual cases must introduce extra variables
  $\Gamma_1',\ldots$ which are later constrained such that e.g. either
  $\Gamma_1' = \Gamma_1,A$ or $\Gamma_1' = \Gamma_1$. This doubling of
  the number of variables in play makes the proofs harder to follow.

  One of the aims of mechanisation is to ensure that the proofs are
  impeccable. Existing presentations are deficient in this regard.
  For example, Girard's presentation contains numerous typographical
  mistakes. Perhaps more worryingly, Girard dismisses the structural
  cases as trivial, and omits the proofs. However, our experience was
  that the structural rules, $\WeakL, \WeakR$,
  combined with sequents that are (pairs of) \emph{sets} of formulae,
  were the \emph{hardest} to get right. We hope their inclusion here
  will clarify what otherwise might have remained a murky part of the
  proof. Certainly we have addressed all relevant cases, so that our
  presentation is complete.
  
  Correctness of the proofs ultimately rests on the foundation
  of the theorem prover in which the mechanisation has taken place.
  Isabelle/HOL is a fully expansive theorem prover, whose kernel is
  small and has often been certified by experts. It is extremely
  unlikely that Isabelle would incorrectly assert that a theorem had
  been proven.
  
  For correctness, one also requires that the definitions correspond
  to the related informal notions.
  We have tried to ensure that this is the case in two ways. We have
  used concrete mathematical structures which directly correspond to
  the intuitive notions wherever possible, rather than resorting to
  sophisticated techniques such as HOAS. Our derivations are concrete
  objects. Our sets of formulae are indeed sets. We have provided a
  standard presentation of first order formulae based on de Bruijn
  indices.  Since much other work has been conducted with de Bruijn
  indices, they are fairly well understood, so that it should be easy
  to convince oneself of the correctness of our concrete presentation.
  On the other hand, we do not want our definitions to be over
  concrete, and so introduce unnecessary complexities. For example, we
  do not want our proofs to take advantage of properties that are
  present only for one particular implementation of formulae. For this
  reason, we have also isolated the weakest possible properties
  required in our proofs. For example, our axiomatic presentation of
  first order formulae, which involves variable binding, is extremely
  weak.  For our particular proof of Craig's Interpolation Theorem,
  these properties cannot be made weaker. These properties should be
  satisfied by any reasonable concrete implementation of first order
  formulae\footnote{This is not quite true, since in order to make the
    mechanisation as slick as possible, we have used equality rather
    than alpha equivalence. A mechanisation based on named variables
    and alpha equivalence would have to quotient the type of formulae
    by alpha in order to satisfy our axioms.}.
  Of course, we tied the two presentations of formulae together by
  proving that the axiomatic properties we require are satisfied by
  the de Bruijn representation.
  
  It is still the case that the informal presentation in this paper,
  which is written by hand, may contain typos and other errors. Until
  the mechanised text becomes primary, this is inevitable. We have
  attempted to prevent errors creeping in by explicitly quoting the
  formal witnesses in the informal text. However, errors may still
  arise. The mechanised text does not suffer from these problems.
  Against this, even our informal presentation surely contains less
  errors and typos than appear in standard presentations. We hope that
  our presentation becomes \emph{definitive}.

\item \textbf{Appropriate use of Automation} To formally prove Craig's
  Interpolation Theorem without automation would be a very lengthy
  task. We have used automation extensively to keep the formal
  mechanised proof to a small size. On the other hand, the only parts
  of the tactic script that are really essential are the initial use
  of induction over the size of the derivation, and the witnesses used
  to instantiate quantifiers. Thus, the proof could be made
  considerably smaller, i.e. the proof could simply be a call to
  automation with the existential witnesses supplied as a hint.
  However, we also wanted to preserve the structure of the proof, so
  that although the proof could be automated in one or two lines of
  tactic script, we prefer to sketch out the main case splits and
  match reasonably high-level subgoals to tactic lines in the
  mechanised proof.
  
\item \textbf{Elegance, Simplicity} Our mechanisation is succinct. Our
  proofs are the weakest and most direct that we could manage.
  Usually there is some trade off in this area. Weakest proofs are
  typically those arrived at using {\Cut} free proofs, and minimal
  strengthening of induction statements. However, it is sometimes the
  case that one can strengthen the induction statement in many ways,
  perhaps so that it is much stronger than required, but such that it
  is syntactically simpler than the minimal strengthening. The only
  possible place where we have strengthened an induction statement is
  in the statement of the strong form of Craig's Interpolation
  Theorem, and this is a standard strengthening which we felt it would
  be unwise to deviate from.  Moreover, we did not see much scope for
  a syntactically simpler version. Other than this, our proofs are
  {\Cut} free, and as weak as they can be. This is what gives rise to
  the very weak axiomatisation of the properties of first order
  formulae.
  
  For us, elegance is closely tied to syntactic properties of proofs
  and definitions. Thus, {\Cut} free proofs are inherently elegant
  because, for example, they proceed without detour, direct from
  assumptions to conclusions. In addition, we have strived to keep our
  definitions simple and elementary. Simplicity aids
  understanding. Our aim in this is that the reader should never at
  any point feel that the development is not completely
  straightforward and elementary. 
  
  As an example of how we achieve simplicity, what is not so obvious
  from the informal and formal mechanised presentation of the result
  is the extent to which we have played around with various
  definitions to allow the mechanisation to be as clear and
  straightforward as possible. For example, the two {\Weak}
  rules are actually admissible. However, since they are used
  extensively to form the derivation witnesses required in the proof,
  we would have to prove them admissible if we omitted them from our
  basic system. This in turn would involve a separate inductive proof
  to show the well known substitutivity property of eigenvariables in
  proofs. This would be a considerable detour, whilst we prefer our
  mechanisation to remain focused solely on the proof of Craig's
  Interpolation Theorem. Craig's Interpolation Theorem is essentially
  a structural theorem, so a detour into eigenvariable properties
  would be out of place and detract from the essence of the proof. For
  these reasons, we include the {\Weak} rules explicitly. The
  cost is that we must treat these cases in the proof.  However, these
  cases are intrinsically interesting, as the hardest cases, and are
  required in other presentations, so that including these cases
  explicitly is a double gain.
  
\item \textbf{Modularity} 
  
  In order to support different implementations of formulae, in this
  particular case the axiomatic version and the version based on de
  Bruijn notation, we have modularised our development. This consists
  of two related tasks.

  \begin{itemize}
  \item Identifying the weakest properties of formulae that are
    required in the main proof of Craig's Interpolation Theorem.
    
  \item Identifying a minimal common language that all implementations
    of formulae have.
  \end{itemize}
  
  To find the weakest properties of formulae, one typically develops a
  {\Cut} free proof, and examines the leaves of the proof to identify
  those that are provable solely in the language of formulae. To
  identify a minimal common language one examines the formulae
  constructs that appear in the main proof and tries to eliminate as
  much as possible.
  
  In fact, these two activities are linked: one cannot conduct a
  proof, or even state the theorem, without some notion of what a
  formula is. On the other hand, the statement of the theorem may
  involve references to formulae constructs that are really redundant,
  yet their presence in the theorem statement forces their use
  throughout the proof.
  
  For example, the notion of substitution which appears in the de
  Bruijn presentation, is present in some form in all concrete
  representations. It is therefore part of the common language.
  However, its absence from our axiomatic presentation of formulae
  indicates that it is not a necessary notion in order to prove
  Craig's Interpolation Theorem. Whilst conducting early versions of
  the proofs, we began to suspect that substitution could be
  eliminated from the common language we were using for formulae, and
  we worked to bring this about. This is related to our previous
  comments on including the weakening rules explicitly.
  
  We discuss these issues further in the section on applications of Craig's
  Interpolation Theorem, Sect. \ref{sectCraigApplications}.

\end{itemize}

%%%%%%%%%%%%%%%%%%%%%%%%%%%%%%%%%%%%%%%%%%%%%%%%%%%%%%%%%%%%%%%%%%%%%%%%%%%%%%%%
\tsection{Applications}

\label{sectCraigApplications}

In the introduction we claimed that Craig's Interpolation Theorem has
many applications. In fact, it is the kind of result that becomes part
of one's way of thinking about mechanisation, such are its diverse
applications. 

Let us immediately repeat that Craig's Interpolation Theorem has a
constructive proof, which is to say, it is an algorithm that
transforms proofs and furnishes the interpolation formula. We have not
expressed it as a deterministic algorithm, because the proof is
essentially non-deterministic, so that determinising it would be
inelegant. Never-the-less it would be simple to write a primitive
recursive function which produced the interpolant.

As another example, Craig's Interpolation Theorem can be used in
automatic proof search. Suppose we have two (disjoint) languages (set
of predicates which may appear in formulae) $L_1,L_2$. We wish to
prove

\[
\seq{\Gamma_1,\Gamma_2}{\Delta_1,\Delta_2}
\]

where $\Gamma_i,\Delta_i$ is expressed in language $L_i$.
By the strengthened interpolation theorem, Thm. \ref{thmCraigStrong}, we can find a formula $C$ such that

\[
\seq{\Gamma_1}{\Delta_1,C} \skipand \seq{C,\Gamma_2}{\Delta_2}
\]

and moreover such that all predicates appearing in $C$ appear also in
$\Gamma_1,\Delta_1$ and in $\Gamma_2,\Delta_2$. But since $L_1,L_2$
are disjoint, $C$ can only be $\bot$ or $\top$\footnote{Or
  conjunctions, disjunctions, negations of $\top,\bot$\ldots which can
  be simplified to $\bot$ or $\top$.}. So

\[
\seq{\Gamma_1}{\Delta_1} {\fskip \textit{or} \fskip} \seq{\Gamma_2}{\Delta_2}
\]

We can then call our automation separately and in parallel on these
two subproblems. In this way we have reduced the search space
considerably, with nothing but syntactic considerations. This is an
example of ``purity of methods''. As another example of purity of
methods, if a sequent expressed in the language of $L_1$ is provable,
it is provable without taking a detour via $L_2$, which is direct from
the subformula property of {\Cut} free derivations. Clearly this is
extremely useful when restricting automation which would otherwise
wander off into the extensive libraries of modern theorem provers in
its search for a proof of some specific statement in a clearly defined
sublanguage.

Let us now consider a more subtle use of Craig's Interpolation
Theorem. Suppose we wish to conduct a mechanisation that uses some
form of variable abstraction and binding. For example, in our
mechanisation we wish to have a representation of first order
formulae. We might wish to use our favourite representation, say, de
Bruijn. The basic type of abstraction provided by de Bruijn
representations binds the free $0$th variable, as is evident in the
datatype of de Bruijn formulae.

\isab{
\isanewline
\isacommand{datatype}\ form\ {\isacharequal} \isanewline
\ldots \isanewline
\ \ {\isacharbar}\ FAll\ form\isanewline
\ \ {\isacharbar}\ FEx\ form\isanewline
}

Whilst this may make sense for a representation where variables are
numbers, it makes no sense for a named representation say, where there
is no inherent notion of order on the variables. 

Craig's Interpolation Theorem suggests that we should pay close
attention to the language we use to state theorems. For example, let
$F$ represent the axioms for our representation of formulae, and $T$
the main theorem we wish to prove. Then we can find a $C$ such that

\[
\seq{F}{C} \skipand \seq{C}{T}
\]

This $C$ is the interface between the subtheory generated by $F$ and
our main theory in which we prove $T$. If we now replace $F$ with some
other implementation of formulae, $F'$, we would have to rephrase $T$
in terms of this new implementation as $T'$, the lemmas $C$ exported
by our theory of formulae would change to $C'$, and much additional
reworking of proofs would result.  To remedy this, we should express
$T$ using formulae constructs that are found in every implementation.
In this case, Craig's Interpolation Theorem assures us that $C$ must
be expressed also in this shared language, and the only rework that is
required when changing formulae representations is in the proof of
$\seq{F'}{C}$.

Returning to our example, if we used the de Bruijn representation directly, our
phrasing of Craig's Interpolation Theorem would include de Bruijn
constructs, and our mechanisation would include much that was specific
to de Bruijn representations.

For this reason, we avoid the basic de Bruijn abstraction, and work
instead with a named abstraction, even though named abstraction is not
a given for our underlying de Bruijn implementation. We do not unnecessarily bias our development towards named implementations either-- rather than instantiate a quantifier $\forall x. A$ as $[t/x]A$, we have an operation of ``instantiation on the top most quantifier'', $\textit{FAll\_inst}\ t\ (\forall x. A)$. Our axiomatic presentation (which is nothing more than the
separate clauses of the interpolant $C$) certainly hides the de Bruijn
specific constructs. The advantage is that we could later substitute
some new implementation of formulae (named, bound/free) without any
additional work in the main theory, though we would of course have to
prove our axioms (the clauses of our interpolant $C$) were satisfied
in this new implementation. In this way we have used Craig's
Interpolation Theorem in the mechanisation of the proof of the theorem
itself!

This approach can also be used to refactor existing theories, since
Craig's Interpolation Theorem transforms existing proofs. 

In my thesis \cite{ridge05thesis} I suggest other ways in which Craig's
Interpolation Theorem can shape a mechanisation.

%%%%%%%%%%%%%%%%%%%%%%%%%%%%%%%%%%%%%%%%%%%%%%%%%%%%%%%%%%%%%%%%%%%%%%%%%%%%%%%%
\tsection{Conclusion}

We presented the first complete mechanisation of Craig's Interpolation
Theorem. We also talked about some aspects of the mechanisation, and
some of the applications of the theorem to mechanised reasoning. In
the main text, we have indicated where the contributions of the paper
lie, and we briefly recap some of these here.

\begin{itemize}
\item Clear, correct and complete formal presentation of Craig's Interpolation Theorem.
\item We have worked hard to isolate the minimal properties we require
  during the proof. For example, we present a very weak
  axiomatisation of first order formulae. For another example, we
  phrase the logical system in such a way that we avoid a detour
  through the eigenvariable properties of derivations. 
\item Complete rendition of mechanised version, save that some proof scripts have been omitted.
\item Discussion of the application of Craig's Interpolation Theorem to mechanisation and automation. 
\item Particularly, we described our development of first order
  formulae with their notion of binding, and how we obtained such a
  weak axiomatisation.
\end{itemize}

There is some related work. In \cite{boulme96craigCoq}, the author
develops a partial proof of Craig's Interpolation Theorem in Coq. This
is based on a single propositional connective, NAND. As the author
admits, the intent was to extend the work to the usual formulae, but
unfortunately this was never attempted. This work is certainly
considerably more involved than that presented here. Moreover, the
importance of these results usually does not lie in the result itself,
but in the details of the proof: if one understands the details, one
can adapt the proof and use variants of the result in one's own work.
Thus, the restriction to a rather unusual connective is indeed a real
restriction, since one has to work much harder to translate the usual
formulae one meets during proof into NAND form. Furthermore, the proof
is sufficiently complicated that much of the beauty of Craig's
Interpolation Theorem has been lost in the details of formalisation.

Further afield, there is much formalised proof theory. Let us briefly
mention the work of Pfenning on formalised {\Cut} elimination
\cite{Pfenning00ic} in Twelf, which is inspirational. A more
sophisticated development is that of strong normalisation for System F
by Altenkirch in LEGO \cite{alti:tlca93}.

%\appendix

\bibliographystyle{alpha}
\bibliography{../root.bib}

\end{document}